\documentclass[10pt,a4paper,twocolumn,english,prb,aps,showpacs,floatfix,groupedaddress,superscriptaddress]{revtex4-1}
\usepackage{graphicx}
\usepackage{epsfig}
\usepackage[english]{babel}
\usepackage{amsmath}
\usepackage{amssymb}
\usepackage{amsfonts}
\usepackage{wasysym}%
\usepackage{longtable}
\usepackage{dcolumn}% Align table columns on decimal point
\usepackage{bm}
\usepackage{bbm}
\usepackage{nicefrac}
\usepackage{color,array}
\usepackage{colortbl}
\usepackage{dsfont}
\usepackage{mathtools}
\usepackage{subfigure}  % use for side-by-side figures
\usepackage[breaklinks,colorlinks=true,citecolor=blue,linkcolor=blue]{hyperref}
\usepackage{breakurl}
\usepackage{soul}

\def\bra#1{\mathinner{\langle{#1}|}}           %bra-state
\def\ket#1{\mathinner{|{#1}\rangle}}           %ket-state

\begin{document}

\newcommand{\be}{\begin{equation}}
\newcommand{\ee}{\end{equation}}
\newcommand{\bearr}{\begin{eqnarray}}
\newcommand{\eearr}{\end{eqnarray}}
\newcommand{\bsubeq}{\begin{subequations}}
\newcommand{\esubeq}{\end{subequations}}
\newcommand{\nn}{\nonumber}
\newcommand{\reqn}{\eqref}
\newcommand{\NN}{n.\@n.\@}

\definecolor{red}{RGB}{220,20,60}
\definecolor{green}{RGB}{0,100, 0}

\def\tcr#1{\textcolor{red}{#1}}
\def\tcb#1{\textcolor{blue}{#1}}
\def\tcg#1{\textcolor{green}{#1}}

\title{From Gapped Excitons to Gapless Triplons in One Dimension}

\author{M. Hafez Torbati}
\email{mohsen.hafez@tu-dortmund.de}
\author{Nils A. Drescher}
\email{nils.drescher@tu-dortmund.de}
\author{G\"{o}tz S. Uhrig}
\email{goetz.uhrig@tu-dortmund.de}

\affiliation{Lehrstuhl f\"ur Theoretische Physik I, Technische Universit\"at Dortmund,
Otto-Hahn-Stra\ss e 4, 44221 Dortmund, Germany}

\date{\rm\today}

\begin{abstract}
Often, exotic phases appear in the phase diagrams between conventional phases. 
Their elementary excitations are of particular interest. Here, we consider the example
of the ionic Hubbard model in one dimension. This model is a band insulator (BI) for
weak interaction and a Mott insulator (MI) for strong interaction. Inbetween,
a spontaneously dimerized insulator (SDI) occurs which is governed by energetically 
low-lying charge and spin degrees of freedom. Applying a systematically controlled
version of the continuous unitary transformations (CUTs) we are able to
determine the dispersions of the elementary charge and spin excitations
and of their most relevant bound states on equal footing. The key idea is
to start from an externally dimerized system using the relative 
weak interdimer coupling as small expansion parameter which finally is set to unity
to recover the original model.
\end{abstract}

% 71.10.Fd 	Lattice fermion models (Hubbard model, etc.)
% 71.10.Li 	Excited states and pairing interactions in model systems
% 71.30.+h 	Metal-insulator transitions and other electronic transitions
% 74.20.Fg 	BCS theory and its development


\maketitle

% \tableofcontents

%%
%%%%%%%%%%%%%%%%%%%%%%%%%%%%%%%%%%%%%%%%%%%%%%%%%%%%%%%%%%%%%%%%%%%%%%%%%%%%%%%%
%%%%%%%%%%%%%%%%%%%%%%%%%%%%%%%%%%%%%%%%
\section{Introduction}
\label{sec:introduction}

In condensed matter physics, an important focus is the understanding
of the phases which determine the qualitative behavior of the systems under study.
In strongly correlated systems, in particular, a large variety of phases
may occur. In order to understand their physical properties their static and dynamic
correlations have to be described. We focus here on the dynamic correlations
of strongly interacting electronic
systems which display insulating behavior in certain parameter regimes.
Our key issue is to understand and to describe the elementary excitations, 
also called quasiparticles, and possible bound states formed by them.

At commensurate fillings of electronic bands the 
band insulator (BI) and the Mott insulator (MI) represent two distinct
classes of insulating systems which exhibit distinct excitations. In a 
BI, the spin and charge gaps are both finite and equal. These finite
gaps in the charge and in the spin channel originate from the electron-ion
interactions~\cite{Gebhard97}. The interaction among the electrons 
may be arbitrarily weak. The elementary excitations in a BI are 
electrons or holes. A pair of an electron  and a hole
may form non-magnetic singlet and/or magnetic triplet bound state(s)
due to the attractive two-particle  interaction.
 
A MI is characterized by  spin and charge excitations which display 
different dispersions. The MI phase is stabilized by a strong electron-electron interaction~\cite{Gebhard97,Imada98}. 
There is no MI without sufficiently strong interaction between the charges.
In the MI, single charges (electron or hole quasiparticles) are gapped with
a significant gap of the order of the interaction. For large repulsive
interactions, the magnetic excitations are described by Heisenberg models
with antiferromagnetic exchange. The precise properties of the
magnetic excitations strongly depend on further details of the underlying
lattice. Both gapped and gaples magnetic excitations may occur.

In one dimension, the elementary spin excitations 
in MIs are established to be spinons which carry the total spin $S=1/2$ and show
a gapless linear dispersion at the edge of the Brillouin zone (BZ)~\cite{Cloizeaux62,fadde81}. 
For any infinitesimal dimerization confinement occurs: Bound states of two spinon with total
spin $S=1$, called triplon, occur and constitute the elementary 
excitations\cite{cross79,uhrig99a,Knetter00,zheng01b,Schmidt03,papen03}.
In higher dimensions, the convential scenario is the occurrence of phases with long range
magnetic order. The generic excitations are magnons, i.e., gapless Goldstone bosons 
with integer spin\cite{auerb94}. But in case of strongly competing interactions, for instance
if three spins should align mutually antiparallel to satisfy their interactions,
far more complex behavior may occur. In particular in two dimensions,
gapped excitations with anyonic statistics may occur \cite{balen10}. 

A general motif for exotic phases to occur are competing driving forces:
If there is a control parameter of the system which implies 
that the system is in the conventional phase A in one limit and in phase
B in the other, it is promising to look closely what happens
at the transition from A to B. Often, a third  phase C occurs
in which the main driving forces essentially cancel, leaving room for
novel ordering mechanisms. Following this spirit, we study
the competition of two driving forces acting directly on the charges.
The two competing, rather conventional phases are the BI and the MI.
The control parameter is the ratio between an on-site repulsion $U$ and
an alternating local potential $\delta$.

The simplest model with these antagonists is the  ionic Hubbard model (IHM). 
The model consists of the usual Hubbard model plus
a staggered ionic potential $\delta$ which splits the energy on even and odd sites.
Its Hamiltonian in 1D is given by
\bearr
H &=& \frac{\delta}{2} \sum_{i,\sigma} (-1)^i n^{\phantom{\dagger}}_{i,\sigma}
+U\sum_{i} \left( n^{\phantom{\dagger}}_{i,\uparrow} -\frac{1}{2} \right) 
\left( n^{\phantom{\dagger}}_{i,\downarrow} -\frac{1}{2} \right) \nn \\
&&+ t \sum_{i,\sigma} \left( c^{\dagger}_{i,\sigma} c^{\phantom{\dagger}}_{i+1,\sigma}
+ {\rm H.c.} \right),
\label{eq:IHM}
\eearr
where the operators $c^{\dagger}_{i,\sigma}$ and $c^{\phantom{\dagger}}_{i,\sigma}$
are the  fermionic operators creating and annihilating an electron with spin 
$\sigma$ at site $i$. 
The operator $n^{\phantom{\dagger}}_{i,\sigma} := c^{\dagger}_{i,\sigma}c^{\phantom{\dagger}}_{i,\sigma}$
counts the number of spin-$\sigma$ electrons at site $i$.

Clearly, the MI phase is stable if the Hubbard interaction is the dominant
term in the system while the BI becomes the ground state if
the ionic potential prevails over the other model parameters.
It is shown that the MI and the BI are separated by an intermediate phase 
in 1D which is known to be a spontaneously dimerized 
insulator~(SDI)~\cite{Fabrizio99, Torio01, Manmana04, Otsuka2005, Tincani09,HafezTorbati2014}.
In two dimensions, however, the nature of the middle phase is highly
disputed~\cite{Garg06,Kancharla07,Paris07,Craco08,Chen10}.

The IHM was first proposed~\cite{Strebel1970,Soos1978,Nagaosa86a} to describe the neutral-ionic transition
in the charge-transfer mixed-stack organic compounds like TTF-choloranil~\cite{Torrance81a}.
Later, it was shown that the model is also a candidate to explain the
ferroelectricity in transition metal oxides such as BaTiO$_3$~\cite{Egami93}. 
Transition metal oxides and mixed-stack organic
compounds~\cite{Kobayashi12} are both interesting classes of solids
which supports the relevance of the IHM beyond its theoretical significance. 

In the present article, we explore the excitation spectrum of the one-dimensional IHM
at zero temperature and half-filling using 
{\it directly evaluated enhanced perturbative continuous unitary
transformations} (deepCUT)~\cite{Krull12}. The underlying idea is to map
the microscopic Hamiltonian \eqref{eq:IHM} to an effective model 
expressed directly in the elmentary excitations. This mapping is systematically
controlled by some small parameter. In previous work, we started
from the BI in which the elementary excitations are unbound, but dressed fermions:
Quasiparticles and quasiholes \cite{Hafez10b,Hafez11,HafezTorbati2014}. The resulting effective model 
can be used within the BI phase and to some extent in the adjacent SDI phase.
But the dispersion in the SDI phase and the MI phase cannot be treated.
For this reason, we take a different view point in the present work.

We start from the dimer limit where the system is composed
of isolated dimers~\cite{Oitmaa06, Duffe11} and turn on the interdimer hopping in
the renormalization scheme of the deepCUT~\cite{Krull12}. The advantage of the dimer
limit, compared to the BI limit~\cite{HafezTorbati2014}, is that it allows us to 
access all the three different phases of the IHM. 
We look for the transition points $U_{c1}$ between the BI and the SDI phase 
and  $U_{c2}$ between the SDI and the MI phase by analyzing the
ground state energy and various energy gaps. The results are compared to data
from a density matrix renormalization (DMRG) calculation~\cite{Tincani09}. The main focus, however,
will lie on the momentum dependent 
low-energy  spectrum  in the three phases: BI, SDI, and MI.

In the BI phase, we verify that the approach from the dimer limit 
satisfactorily reproduces the deepCUT results obtained from the BI 
limit~\cite{HafezTorbati2014}. 
In the intermediate SDI phase, both charge and spin degrees 
of freedom contribute to the low-energy spectrum. We discuss the 
difficulties of the electron-hole picture to explain this excitation
spectrum.
In the MI phase, it is found that
the low-energy physics of the IHM for large enough Hubbard interaction 
can be described by an effective Hamiltonian {\it purely} in 
terms of magnetic triplon operators in the spirit of the description
of spin chains from the dimer limit in Ref.~\onlinecite{Schmidt03}.
The analysis of the ensuing effective triplon Hamiltonian 
yields quantitative results for the gapless triplon dispersion of the IHM in
the MI phase.

The article is set up in the following way. After this Introduction,
we discuss the dimer limit and its local excitations in detail in 
Sect.\ \ref{sec:dimer_limit}. Section \ref{sec:technical_aspects} is devoted to 
a brief discussion of technical aspects.
Next, we present results for the ground state energy and the gaps
to the lowest excitations in Sect.\ \ref{sec:gaps}. In Sect.\ 
\ref{sec:spectrum} we elucidate the dispersions, i.e., the full
momentum dependence of the elementary excitations in the BI, in the SDI,
and in the MI phase. Finally, the article is concluded in Sect.\ 
\ref{sec:conclusion}.

\section{Dimer Limit}
\label{sec:dimer_limit}

The IHM~\reqn{eq:IHM} has a four dimensional Hilbert space at each site: The empty 
state, the spin up and down states, and the  doubly occupied state. However, the empty states 
on odd sites and the doubly occupied states on even sites lie very high in energy
for $U,\delta \gg t$. In the following, we limit our analysis of the IHM to the 
case where $U,\delta \gg t$ and truncate the Hilbert space such that no empty state
on odd sites and no doubly occupied state on even sites is considered. 
We stress that such a restriction 
has no qualitative effect on the phase diagram of the IHM for small hopping $t$. 

In addition, a DMRG study shows that the position of 
the transition points of the IHM with the truncated Hilbert space quite accurately match
the results of the  IHM considering the full Hilbert space in the limit $U,\delta \gg t$.
For instance, $t=\delta/20$, $U_{c1}=1.065\delta$  in the truncated case \cite{Tincani09} and 
$U_{c1}=1.069\delta$ in the untruncated case \cite{Manmana04}.
Of course, one could apply a first CUT to eliminate the high energy states as we
did in previous work \cite{HafezTorbati2014}. In view of the minute difference in number we
refrain from this first step in order to keep the initial Hamiltonian as simple
as possible. But we stress that for larger ratios $t/\delta$ such a first step
is indicated.

We will treat the odd and even sites in the same way by restoring
the translational invariance by an electron-hole transformation on the odd sites
\be
c^{\dagger}_{i,\sigma} \rightarrow \eta^{\phantom{+}}_\sigma h^{\phantom{\dagger}}_{i,\bar{\sigma}},
\ee
where $\eta^{\phantom{+}}_\uparrow=1$, $\eta^{\phantom{+}}_\downarrow=-1$, and 
$\bar{\sigma}$ stands for the opposite direction of the spin.
Unifying the electron and hole operators by the fermion operator
\be
f_{i,\sigma} := 
\begin{cases}
c^{\phantom{\dagger}}_{i,\sigma} & {\rm for} ~ i \in {\rm even}, \\
h^{\phantom{\dagger}}_{i,\sigma} & {\rm for} ~ i \in {\rm odd},
\end{cases}
\label{eq:fermion_op}
\ee
maps the Hamiltonian~\reqn{eq:IHM} to the form
\bearr
H &=& \frac{U\!-\!2\delta}{4} \sum_i \mathds{1} 
+ t \sum_{i,\sigma} \eta^{\phantom{+}}_\sigma ( f^\dagger_{i,\sigma} f^\dagger_{i+1,\bar{\sigma}} + {\rm H.c.} ) \nn \\
&&+ \frac{\delta - U}{2} \sum_{i,\sigma} f^\dagger_{i,\sigma} f_{i,\sigma}^{\phantom{\dagger}}
+ U \sum_i f^\dagger_{i,\uparrow}f^\dagger_{i,\downarrow} 
f_{i,\downarrow}^{\phantom{\dagger}}f_{i,\uparrow}^{\phantom{\dagger}}.
\label{eq:IHM_frep}
\eearr
In this operator representation, omitting all the doubly occupied states is equivalent to 
omitting the empty states on the odd sites and the doubly occupied states on the even sites
in the original picture \eqref{eq:IHM} as we intended to do. 

\begin{figure}[t]
\includegraphics[width=1.0\columnwidth,angle=0]{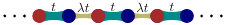}
  \caption{(Color online) Schematic representation of dimer limit expansion. The intradimer 
  hopping parameter is denoted by $t$, the interdimer hopping by $\lambda t$.
  For $\lambda=0$ the system consists of isolated dimers and for $\lambda=1$
  the uniform chain is retrieved.}
  \label{fig:dimer}
\end{figure}

Hubbard operators
are used to decompose the Hamiltonian~\reqn{eq:IHM_frep} into the terms which create \emph{or} annihilate specific numbers of double occupancies. The Hubbard operators are defined by
\begin{subequations}
 \begin{align}
g^\dagger_{i,\sigma} &:=  \ket{\sigma}_i \prescript{}{i}{\bra{e}} , \\
g^\dagger_{i,d} &:=  \ket{d}_i \prescript{}{i}{\bra{e}}, 
\end{align}
\end{subequations}
where $\sigma=\uparrow,\downarrow$ stands for the spin direction while
$d$ and $e$ stand for the doubly occupied state and the empty state, respectively. 
The $f$-operator in terms of Hubbard operators is given by
\be
f^\dagger_{i,\sigma} = g^\dagger_{i,\sigma}+
\eta^{\phantom{+}}_\sigma g^\dagger_{i,d}g^{\phantom{\dagger}}_{i,\bar{\sigma}}.
\label{eq:f2g}
\ee
Finally, inserting Eq.~\reqn{eq:f2g} and its hermitian conjugate into the 
Hamiltonian~\reqn{eq:IHM_frep} and only keeping the terms which matter for
the subspace with zero double occupancies yields
\bearr
\label{eq:IHM_res}
H &=& \frac{U\!-\!2\delta}{4} \sum_i \mathds{1} 
+ \frac{\delta - U}{2} \sum_{i,\sigma} g^\dagger_{i,\sigma} g_{i,\sigma}^{\phantom{\dagger}}  \nn 
\\
&&+ t \sum_{i,\sigma} \eta^{\phantom{+}}_\sigma ( g^\dagger_{i,\sigma} g^\dagger_{i+1,\bar{\sigma}} + {\rm H.c.} ).
\eearr
\begin{figure}[t]
\includegraphics[width=1\columnwidth,angle=0]{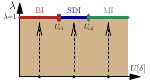}
  \caption{(Color online) Schematic phase diagram of the model~\reqn{eq:IHM_dimer} for a fixed 
  value of the hopping parameter. The system is dimerized for all values $\lambda<1$. At $\lambda=1$
  we reach the phase diagram of the ionic Hubbard model (IHM). By choosing an appropriate value for the Hubbard 
  interaction~$U$ and increasing the parameter $\lambda$ from 0 to 1 we can describe different phases of
  the IHM: the band insulator (BI), the spontaneously dimerized insulator (SDI), and the Mott insulator (MI).}
  \label{fig:dimer_phase}
\end{figure}

In order to use the dimer limit, see Fig.~\ref{fig:dimer}, 
as starting point for the intended deepCUT analysis we
modify the Hamiltonian~\reqn{eq:IHM_res} such that it takes the form
\bearr
H &=& \frac{U\!-\!2\delta}{4} \sum_i \mathds{1} 
+ \frac{\delta - U}{2} \sum_{i,\sigma} g^\dagger_{i,\sigma} g_{i,\sigma}^{\phantom{\dagger}} \nn \\
&&+ t \!\!\!\! \sum_{i \in {\rm even},\sigma} \!\!\!\!\! \eta^{\phantom{+}}_\sigma ( g^\dagger_{i,\sigma} g^\dagger_{i+1,\bar{\sigma}} 
+\lambda g^\dagger_{i+1,\sigma} g^\dagger_{i+2,\bar{\sigma}} + {\rm H.c.} ), 
\label{eq:IHM_dimer}
\eearr
where $\lambda$ is the perturbative parameter on which we base the truncation of the flow equations 
in the deepCUT~\cite{Krull12}. For $\lambda=0$ the Hamiltonian is composed of isolated dimers and 
for $\lambda=1$ the uniform IHM~\reqn{eq:IHM_res} is retrieved. By fixing an appropriate value for
the Hubbard interaction~$U$ and increasing the parameter $\lambda$ from 0 to 1 we access the three
different phases of the IHM, see Fig.~\ref{fig:dimer_phase}. One notices that the BI and the MI phases
are both on the border of the phase diagram.

For applying the deepCUT method, we re-express
the Hamiltonian~\reqn{eq:IHM_dimer} in terms of 
creation and annihilation operators of the elementary excitations
on a dimer. For vanishing interdimer hopping, $\lambda=0$, the system consists of independent
dimers with a nine dimensional local Hilbert space. The 
nine eigenstates and eigenvalues of a single dimer are summarized in Table~\ref{tab:dimer}.
The ground state energy $\epsilon_0$, the triplon energy $\epsilon_t$, the fermion energy $\epsilon_f$, and
the singlon energy $\epsilon_s$ are found to be
\begin{subequations}
\label{eq:local_energy}
\begin{align}
\epsilon_0 &=-\frac{1}{2} \left( \delta + \sqrt{ (U-\delta)^2+8t^2 } \right), 
\label{eq:local_energy_0} \allowdisplaybreaks[4] \\
\epsilon_t &=+\frac{1}{2} \left( \delta-U + \sqrt{ (U-\delta)^2+8t^2 } \right), 
\label{eq:local_energy_t} \allowdisplaybreaks[4] \\
\epsilon_s &= 2\epsilon_f= \sqrt{ (U-\delta)^2+8t^2 }. 
\label{eq:local_energy_s}
\end{align}
\end{subequations}
The coefficients $\alpha$ and $\beta$ are given by
\begin{subequations}
\label{eq:coeff}
\begin{align}
\alpha &= \sqrt{\frac{1}{2}+\frac{U-\delta}{4\epsilon_f}}, \allowdisplaybreaks[4] \\
\beta &= \sqrt{\frac{1}{2}-\frac{U-\delta}{4\epsilon_f}}. 
\end{align}
\end{subequations}
For all values of the parameters $t$, $U$, and $\delta$, the state with lowest 
energy has total spin zero and it is denoted as vacuum $\ket{0}$. 
There are four degenerate fermionic excited states corresponding to the fermion being
placed on the left site $\ket{f}_{l,\sigma}$ or on the right site $\ket{f}_{r,\sigma}$
and it may take one of two different spin states $\sigma=\uparrow, \downarrow$. 

\begin{table}
\caption{\label{tab:dimer} 
Eigenstates and eigenvalues of a single dimer of
Hamiltonian~\reqn{eq:IHM_dimer}, i.e., at $\lambda=0$. There are three 
different possible states on each site: empty state $e$, spin up state $\uparrow$, and 
spin down $\downarrow$ state. This leads to nine eigenstates on each dimer. 
The ground state has total spin zero and is denoted as vacuum by $\ket{0}$. 
There are four fermionic and four bosonic excited states. The 
expressions $\epsilon_0$, $\epsilon_t$, $\epsilon_f$, and $\epsilon_s$ are defined 
in Eq.~\reqn{eq:local_energy} and the coefficients $\alpha$ and 
$\beta$ are given in Eq.~\reqn{eq:coeff}.
}
\begin{ruledtabular}
\begin{tabular}[b]{ | c | l | c | }
\# & Dimer Eigenstates & Eigenvalues \\ \hline
$1$ & $\ket{0}=-\beta\ket{e,e}
+\frac{\alpha}{\sqrt{2}} \left( \ket{\uparrow,\downarrow}-\ket{\downarrow,\uparrow} \right)$ 
& $\epsilon^{\phantom{+}}_0$ \\
$2$ & $\ket{t}_{+1}=\ket{\uparrow,\uparrow}$ 
& $\epsilon^{\phantom{+}}_0+\epsilon^{\phantom{+}}_t$ \\ 
$3$ & $\ket{t}_{0}=\frac{1}{\sqrt{2}} \left( \ket{\uparrow,\downarrow}+\ket{\downarrow,\uparrow} \right)$ 
& $\epsilon^{\phantom{+}}_0+\epsilon^{\phantom{+}}_t$ \\
$4$ & $\ket{t}_{-1}=\ket{\downarrow,\downarrow}$ 
& $\epsilon^{\phantom{+}}_0+\epsilon^{\phantom{+}}_t$ \\
$5$ & $\ket{f}_{l,\uparrow}=\ket{\uparrow,e}$ 
& $\epsilon^{\phantom{+}}_0+\epsilon^{\phantom{+}}_f$ \\
$6$ & $\ket{f}_{l,\downarrow}=\ket{\downarrow,e}$ 
& $\epsilon^{\phantom{+}}_0+\epsilon^{\phantom{+}}_f$ \\ 
$7$ & $\ket{f}_{r,\uparrow}=\ket{e,\uparrow}$ 
& $\epsilon^{\phantom{+}}_0+\epsilon^{\phantom{+}}_f$ \\ 
$8$ & $\ket{f}_{r,\downarrow}=\ket{e,\downarrow}$ 
& $\epsilon^{\phantom{+}}_0+\epsilon^{\phantom{+}}_f$ \\ 
$9$ & $\ket{s}=+\alpha\ket{e,e}
+\frac{\beta}{\sqrt{2}} \left( \ket{\uparrow,\downarrow}-\ket{\downarrow,\uparrow} \right)$ 
& $\epsilon^{\phantom{+}}_0 + \epsilon^{\phantom{+}}_s$ \\
\end{tabular}
\end{ruledtabular}
\end{table}

Among the four bosonic
excited states, there is one state with total spin zero $\ket{s}$, which we 
call singlon henceforth, and a three-fold degenerate 
triplet with total spin one and  magnetic quantum numbers $\ket{t}_{\pm 1,0}$, which
we call triplon henceforth. It is seen
from Eqs.~\reqn{eq:local_energy_t} and \reqn{eq:local_energy_s} that the local singlon energy $\epsilon_s$ is twice the fermion energy $\epsilon_f$. 
The triplon energy $\epsilon_t$ and the fermion
energy $\epsilon_f$ are close to each other for $U\approx \delta$. 
This makes it difficult to decouple the two-fermion 
sector, the two-triplon sector, and the one-singlon sector from one another
in the deepCUT.

Next, we define the following local hardcore creation operators at the {\it dimer} position 
$j$
\begin{subequations}
\label{eq:dimer_op}
 \begin{align}
 \label{eq:dimer_op_fer}
f^{\dagger}_{j;p,\sigma} &:= \ket{f}_{j;p,\sigma}~\prescript{}{j}{\bra{0}} \quad ; \quad p=l,r 
\allowdisplaybreaks[4] \\
\label{eq:dimer_op_trip}
t^{\dagger}_{j;m} &:= \ket{t}_{j;m}~\prescript{}{j}{\bra{0}} \quad ; \quad m=\pm1,0 
\allowdisplaybreaks[4] \\
\label{eq:dimer_op_sing}
s^{\dagger}_{j} &:= \ket{s}_{j}~\prescript{}{j}{\bra{0}}.
\end{align}
\end{subequations}
The fermion operator $f^{\dagger}_{j;p,\sigma}$ creates a fermionic excitation from the vacuum
at dimer $j$ with spin $\sigma$ at the internal position $p=l$ or $r$ where $l$
stands for the left position and $r$ for the right one.
Similarly, the triplon operator $t^{\dagger}_{j;m}$ and the singlon operator $s^{\dagger}_{j}$
create a triplon with magnetic number $m$ and a singlon at the dimer position $j$, respectively. 
Summarizing, we call these operators of second quantization ``dimer excitation operators''.

\begin{widetext}
The Hubbard $g$-operators can be expressed in terms of the dimer excitation operators 
\begin{subequations}
\label{eq:g2dimer}
 \begin{align}
g^\dagger_{j;l,\sigma}  =& -\beta f^\dagger_{j;l,\sigma} 
+ t^{\dagger}_{j;\eta^{\phantom{+}}_{\sigma}} f^{\phantom{\dagger}}_{j;r,\sigma}
+\alpha f^{\dagger}_{j;l,\sigma} s^{\phantom{\dagger}}_{j}
% \nn \allowdisplaybreaks[4] \\ 
+ \frac{1}{\sqrt{2}}\left( \eta^{\phantom{+}}_{\sigma} \alpha 
+ t^\dagger_{j;0} + \eta^{\phantom{+}}_{\sigma} \beta s^\dagger_{j} \right) 
f^{\phantom{\dagger}}_{j;r,\bar{\sigma}}, 
\allowdisplaybreaks[4] \\
g^\dagger_{j;r,\sigma}  = &-\beta f^\dagger_{j;r,\sigma} 
- t^{\dagger}_{j;\eta^{\phantom{+}}_{\sigma}} f^{\phantom{\dagger}}_{j;l,\sigma}
+\alpha f^{\dagger}_{j;r,\sigma} s^{\phantom{\dagger}}_{j}
% \nn \allowdisplaybreaks[4] \\ 
+ \frac{1}{\sqrt{2}}\left( \eta^{\phantom{+}}_{\sigma} \alpha 
- t^\dagger_{j;0} + \eta^{\phantom{+}}_{\sigma} \beta s^\dagger_{j} \right) 
f^{\phantom{\dagger}}_{j;l,\bar{\sigma}},
\end{align}
\end{subequations}
where $g^\dagger_{j;l,\sigma}$ and $g^\dagger_{j;r,\sigma}$ act on the left and on the 
right site, respectively, of dimer $j$. Finally, the 
Hamiltonian~\reqn{eq:IHM_dimer} in terms of dimer excitation operators reads
\bearr
H=\epsilon^{\phantom{+}}_0 \sum_j \mathds{1} 
+ \sum_j \sum_{m=\pm1,0} \epsilon^{\phantom{+}}_t t^{\dagger}_{j;m}t^{\phantom{\dagger}}_{j;m}
+ \sum_j \epsilon^{\phantom{+}}_s s^{\dagger}_{j} s^{\phantom{\dagger}}_{j} 
% \nn \allowdisplaybreaks[4] \\
+ \sum_{j,\sigma} \sum_{p=l,r} \epsilon^{\phantom{+}}_f 
f^{\dagger}_{j;p,\sigma}f^{\phantom{\dagger}}_{j;p,\sigma} 
% \nn \allowdisplaybreaks[4] \\
+ \lambda t \sum_{j,\sigma} \eta^{\phantom{+}}_{\sigma} \!\!
\left( g^\dagger_{j;r,\sigma} g^\dagger_{j+1;l,\bar{\sigma}} + {\rm H.c.} \right),
\label{eq:IHM_dimer_rep}
\eearr
where the sum $j$ runs over the dimer
positions instead of the original sites. In the last term, the 
Hubbard $g$-operators is meant to be replaced according to Eq.~\reqn{eq:g2dimer}. 
We do not display the resulting expression explicitly for the sake of brevity.
\end{widetext}

If no interdimer hopping is included, the dimer excitations are the true quasiparticles of the 
system. But for any finite value of the relative interdimer hopping $\lambda$ the dimer
excitations start to propagate in the lattice and become dressed quasiparticles.
To remind the reader of this dressing, we refer to them as singlon and triplon and not as singlet
and triplet.
In the next Sects.~\ref{sec:gaps} and \ref{sec:spectrum}, the Hamiltonian~\reqn{eq:IHM_dimer_rep} 
is mapped continuously to effective Hamiltonians such that the dimer excitations can still be used as
quasiparticles of the system even for $\lambda \neq 0$.

\section{Some Technical Aspects}
\label{sec:technical_aspects}

Continuous unitary transformations (CUTs) or 
the flow equation method~\cite{Wegner94,Kehrein06} 
represents the basis of various perturbative~\cite{Knetter00,Knetter03a,Yang10} 
and renormalization approaches~\cite{Dusuel04,Yang11,Krull12,Fauseweh13}.
The basic flow equation reads
\be
\label{eq:flow}
\partial_\ell H(\ell) = [\eta(\ell),H(\ell)]
\ee
where $\ell$ is an auxiliary parameter which parametrizes
the unitary transformation and changes from $\ell=0$ to $\ell=\infty$.
The important choice is how the Hamiltonian is transformed which amounts up
to the choice of the infinitesimal generator $\eta(\ell)=-\eta(\ell)^\dag$.
Once one has chosen a basis for operators in order
to be able to write general operators as linear combinations, 
the differential equation \eqref{eq:flow} induces differential equations
in the coefficients for the basis operators. Generally, the required number
of basis operators is infinite.

The deepCUT is a renormalizing approach which truncates the contributions to 
these differential equations on the
basis of their order in the expansion parameter. The idea is to
\emph{target} a certain quantity, for instance the ground state energy and perhaps
the dispersions, in a certain order $n$. Then all contributions in the flow equations
which are relevant for the targeted quantities up to order $n$ are kept, but
contributions which matter only in higher orders are neglected.
Finally, the resulting set of differential equations is solved numerically.
Thereby, the data obtained comprises contributions in all orders in the
expansion parameter, but it is still approximate. 

Increasing the targeted order we study whether the results still change
significantly. If the results
do not depend on the order of the calculations they can be considered
reliable. For further  technical
aspects of the deepCUT approach we refer the reader to Ref.~\onlinecite{Krull12}. 

Here we discuss specific aspects of the generator that we employ.
The general aim is to obtain an effective Hamiltonian which conserves
the number of excitations \cite{Knetter00,Fischer10}. To this end, 
a particle-conserving generator is used which consists
of the terms occurring in the Hamiltonian with a relative sign depending
on the change of the number of quasiparticles: Positive for terms
incrementing this number and negative for terms decreasing it.

Because the unperturbed part of the Hamiltonian~\reqn{eq:IHM_dimer_rep} has a 
non-equidistant spectrum, the mere number of excitations is not
a suitable criterion to decide about the sign in the generator.
We use the sign of the change in the local energy instead.
Assume $H^{c_t,c_f,c_s}_{a_t,a_f,a_s}$ stands 
for the part of the Hamiltonian which creates $c_t$ triplons, $c_f$ fermions, and $c_s$ singlons
and annihilates $a_t$ triplons, $a_f$ fermions, and $a_s$ singlons. Then this part
contributes to the generator according to 
\be
\hat{\eta} \left[ H^{c_t,c_f,c_s}_{a_t,a_f,a_s}(\ell) \right] = 
{\rm sign}\left( \Delta\epsilon(\ell) \right)H^{c_t,c_f,c_s}_{a_t,a_f,a_s}(\ell),
\label{eq:generator}
\ee
where the local energy change $\Delta\epsilon(\ell)$ is defined by
\be
\Delta\epsilon(\ell) := (c_t-a_t) \epsilon_t(\ell) + (c_f-a_f)\epsilon_f(\ell) 
+ (c_s-a_s) \epsilon_s(\ell).
\label{eq:energy_change}
\ee
The functions $\epsilon_t(\ell)$, $\epsilon_f(\ell)$, and $\epsilon_s(\ell)$ are the onsite
energies of triplons, fermions, and singlons in the course of the flow $\ell$. Their initial
values at $\ell=0$ are given  in Eq.~\reqn{eq:local_energy}. 

In order to set up the differential equations in high targeted order $n$
certain simplification rules are used. Their
key idea is to identify unnecessary contributions early in the calculations
to reduce the required memory resources and to avoid cumbersome follow-up
computations altogether.
The simplification rules that we used in the dimer limit analysis are explained
in Appendix~\ref{sec:app}.

\section{Ground State Energy and Energy Gaps}
\label{sec:gaps}

\subsection{Ground State Energy}

To determine the ground state energy, the state without any singlon, fermion, or triplon has 
to be decoupled from the remaining Hilbert space. We generalize the definition of the reduced
generator introduced in Ref.~\onlinecite{Fischer10} to the case where there are different
kinds of excitations. The ground state generator 
$\eta^{\phantom{\dagger}}_{t:0;f:0;s:0}$ is given by 
\be
\eta^{\phantom{\dagger}}_{t:0;f:0;s:0}(\ell) 
= \sum_{ijk} \left( \hat{\eta}\left[ H_{0,0,0}^{i,j,k}(\ell) \right] 
- {\rm H.c.} \right),
\label{eq:gs_generator}
\ee
where the superoperator $\hat{\eta}$ is defined in Eq.~\reqn{eq:generator}.
The generator~\reqn{eq:gs_generator} is used to decouple the state with zero 
number of triplons, fermions, and singlons from subspaces with finite numbers of
excitations.
We have been able to reach order $12$ in the relative interdimer hopping $\lambda$ targeting
the ground state energy (GSE) based on the dimer limit. Finally, all data is shown
for $\lambda=1$. 

One may ask what is the maximum range of processes which is 
captured by targeting the ground state at order 12? 
To create a pair of fermions with a distance of $n$ dimers we need 
at least $n$ orders, see Eq.~\reqn{eq:g2dimer}. 
These two fermions can not be canceled individually. We need $n$
additional orders to bring these two fermions to nearest-neighbor (\NN) dimers and 
to cancel them. It can be easily seen from Eq.~\reqn{eq:g2dimer} that dealing with 
bosons (singlon and triplons) is even more costly. Hence, all the
terms with an extension larger than $n/2$ are irrelevant at order $n$ targeting the ground state.
This is a nice example how one can use simplification rules to avoid unnecessary terms.
Therefore, targeting the ground state up to order $12$ involves processes with an 
extension of at most $6$ dimers.  
Since the dimer-dimer distance takes two lattice spacings this corresponds to
12 lattice spacings or more presicely to an extension of 13 lattice spacings since 
7 dimers/14 sites are affected by the term.

For comparison, we also analyzed the restricted IHM~\reqn{eq:IHM_res} in the BI limit 
as we did in Ref.~\onlinecite{HafezTorbati2014} for the IHM with the full Hilbert space, i.e.,
allowing also for empty states on odd sites and double occupancies on even sites.
We stress again that the quantitative differences are very small in the parameter
regime considered. Starting from the BI limit, the GSE of the restricted IHM 
is obtained from a deepCUT up to order $20$ in the hopping parameter $t$. 
This means that processes up to a range of $10$ lattice spacings
are included.

\begin{figure}[tb]
\includegraphics[width=1.0\columnwidth,angle=0]{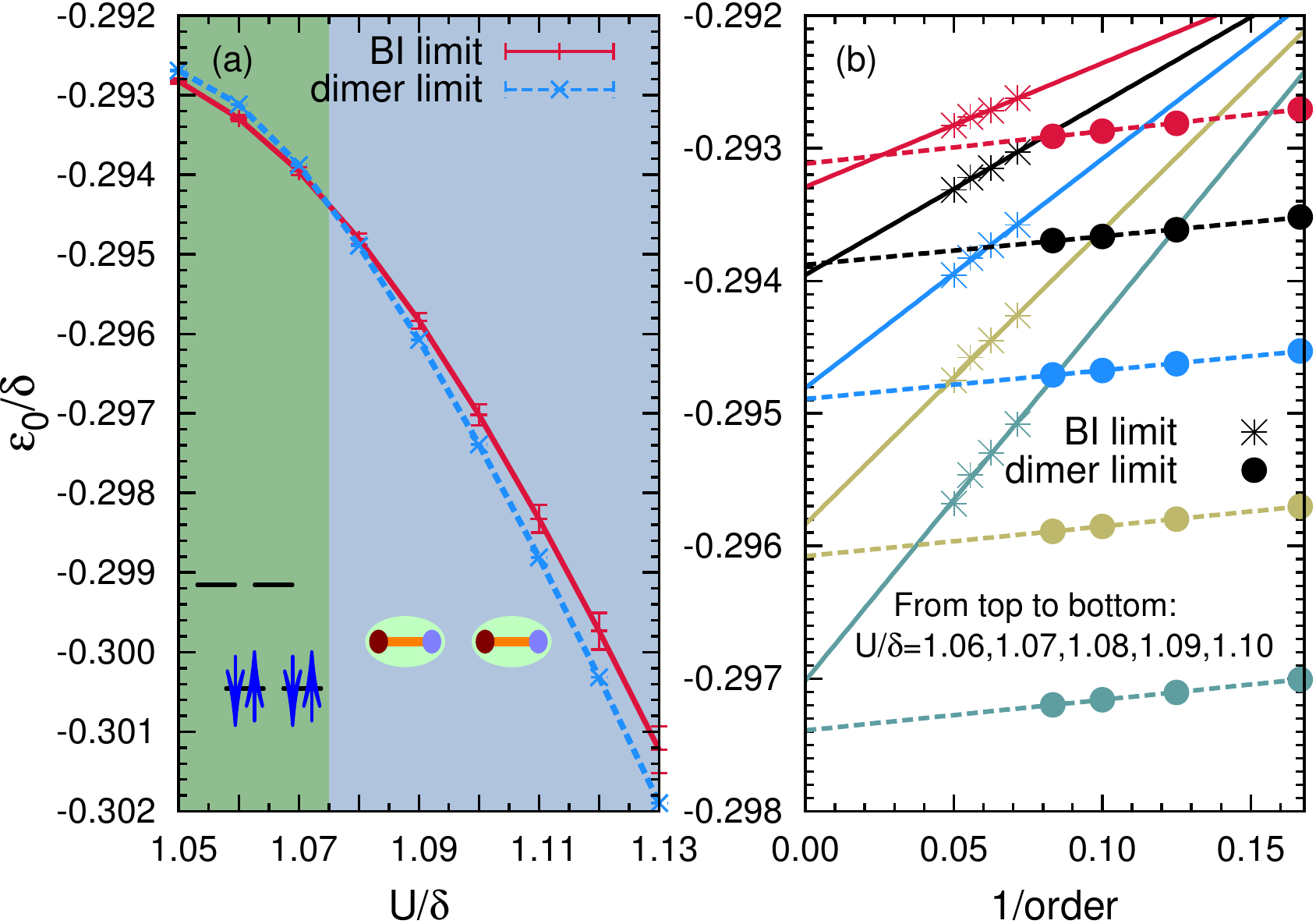}
  \caption{(Color online) Ground state energy per site $\epsilon_0$ of the restricted
  IHM~\reqn{eq:IHM_res} for $t=0.05\delta$. The results
  obtained from both the dimer limit and the BI limit are displayed. 
  In the left panel (a) $\epsilon_0$ extrapolated to infinite
  order is plotted versus the Hubbard interaction $U /\delta$. 
  Beyond the critical point $U_{c1} = 1.075 \delta$, the deepCUT based on the
  dimer limit yields a lower ground state energy than the calculation based on the BI limit.
  In the right panel (b), $\epsilon_0$ is depicted versus the inverse order for 
  various values of $U$. The finite order results are extrapolated linearly to infinite order.}
  \label{fig:GSE}
\end{figure}

The left panel of Fig.~\ref{fig:GSE} depicts the GSE per site extrapolated 
to infinite order versus the Hubbard interaction $U/\delta$ for $t=0.05\delta$. 
The right panel of Fig.~\ref{fig:GSE} shows the extrapolation as a linear fit
in the inverse order. In the right panel, the parameters are $t=0.05\delta$
and  $U=1.06\delta$, $1.07\delta$, $1.08\delta$, $1.09\delta$, and $1.10\delta$
from top to bottom for both the dimer limit (solid circles) and the BI limit (asterisks). 
The dimer limit results exhibit a faster convergence compared to
the results from the BI limit. It is seen in Fig.~\ref{fig:GSE} that the GSE obtained 
from the dimer limit takes lower values than the results from the BI limit
beyond the critical Hubbard interaction $U_{c1}=1.075\delta$ while it is the other way
round for lower $U$. In a rigorous calculation, both results for the GSE 
have to coincide perfectly in the BI phase. But due to the truncations,
 the calculation based on the BI limit works better
in the BI phase and the one based on the dimer limit works better in the SDI phase.
Hence we interprete the results in  Fig.~\ref{fig:GSE} as strong evidence for the 
phase transition BI $\to$ SDI at the intersection $U_{c1}$ of both curves.
For comparison, we state that a DMRG study finds the first transition point 
of the restricted IHM~\reqn{eq:IHM_res} at $U=1.065\delta$~\cite{Tincani09}
so we conclude that our results are exact within about $1\%$.

\subsection{Gaps to Excited States}

Three different gaps can be defined to measure the energy difference between the 
ground state and different excited states.
The charge gap $\Delta_c$ is defined as the energy
needed to add an electron to the system plus the energy to take an electron 
out. It is given by
\begin{subequations}
\be
\Delta_c :=  E_0(N+1)+E_0(N-1)-2E_0(N),
\label{eq:charge_gap}
\ee
where $E_0(N)$ stands for the ground state energy of the system with $N$ 
particles.
The singlet exciton gap $\Delta_e$ and the spin gap $\Delta_s$ are defined as the excitation energy in the channel with the same particle number as the ground state, but with total spin 
zero and one, respectively. They read
\bearr
\Delta_e &:= & E_1(N,S=0)-E_0(N,S=0),
\label{eq:exciton_gap} \\
\Delta_s &:= & E_1(N,S=1)-E_0(N,S=0),
\label{eq:spin_gap}
\eearr
\end{subequations}
where $E_1(N,S)$ denotes the first excited state with $N$ particle and total
spin $S$. 
Because we focus on half-filling, we put $N=L$ where $L$ is the lattice size.

In our formalism, the charge gap $\Delta_c$ can be accessed by decoupling the ground state and
the one-fermion sector (or more sectors) from the other quasiparticle sectors. 
Once the hopping in the one-fermion sector is known
a Fourier transform provides the fermionic dispersion. Twice the minimal energy
of this dispersion yields $\Delta_c$.
The one-fermion generator
$\eta^{\phantom{\dagger}}_{t:0;f:1;s:0}$, which separates the ground state and the 1-fermion
sector from the remaining Hilbert space, is given by
\begin{subequations}
\be
\eta^{\phantom{\dagger}}_{t:0;f:1;s:0}(\ell) = \eta^{\phantom{\dagger}}_{t:0;f:0;s:0}(\ell) 
+\eta^{p}_{t:0;f:1;s:0}(\ell),
\label{eq:f1_generator}
\ee
where we defined 
\be
\eta^{p}_{t:0;f:1;s:0}(\ell) :=  \sum_{ijk} \left( \hat{\eta}\left[ H_{0,1,0}^{i,j,k}(\ell) \right] 
- {\rm H.c.} \right) X^{i,j,k}_{0,0,0},
\label{eq:f1_pgenerator}
\ee
with 
\be
X^{i,j,k}_{i',j',k'} :=  1-\delta_{i,i'} \delta_{j,j'}\delta_{k,k'}.
\label{eq:X_def}
\ee
\end{subequations}
One needs to include the term $X^{i,j,k}_{0,0,0}$ in Eq.~\reqn{eq:f1_pgenerator} 
in order to prevent the interaction $H_{0,1,0}^{0,0,0}$ to appear spuriously twice 
in the generator~\reqn{eq:f1_generator}.

In a similar way, the one-triplon channel can be decoupled
using the one-triplon generator
\begin{subequations}
\be
\eta^{\phantom{\dagger}}_{t:1;f:0;s:0}(\ell) = \eta^{\phantom{\dagger}}_{t:0;f:0;s:0}(\ell) 
+\eta^{p}_{t:1;f:0;s:0}(\ell),
\label{eq:t1_generator}
\ee
with the definition
\be
\eta^{p}_{t:1;f:0;s:0}(\ell) = 
\sum_{ijk} \left( \hat{\eta}\left[ H_{1,0,0}^{i,j,k}(\ell) \right] 
- {\rm H.c.} \right)X^{i,j,k}_{0,0,0}.
\label{eq:t1_pgenerator}
\ee
\end{subequations}
The application of the generator~\reqn{eq:t1_generator} yields an effective Hamiltonian whose
one-triplon sector is separated from the remaining Hilbert space. Thus it is easily
diagonalized by a Fourier transform leading to the 
triplon dispersion. Subsequently, the triplon gap can be found as the minimum of
the triplon dispersion. The triplon gap is an energy gap in the magnetic \mbox{spin-$1$} channel.
Still, one must be cautious to identify the triplon gap with the spin gap because
there may be another $S=1$ excitation with a lower energy.

In the two-fermion sector, there is also a channel with total
spin one and the true spin gap~\reqn{eq:spin_gap} is the minimum of the 
lowest spin-$1$ two-fermion excitation energy and the triplon gap. 
Of course, one could imagine even more sophisticated possibilities for
spin-$1$ excitations, but these two cases are clearly the most probable ones. 
Our results show that the triplon gap is always identical or smaller
than the spin-$1$ two-fermion excitation energy and hence we conclude that the triplon gap 
is indeed the spin gap.

In analogy to the $S=1$ case, there are also two possibilities for the $S=0$  gap. 
Because the singlon quasiparticle
has total spin zero, the $S=0$ gap can be either the singlon gap or the gap in the $S=0$ two-fermion
channel. We find that there is a bound state in the $S=0$ channel of the two-fermion sector
which is significantly lower in energy than the singlon gap. Therefore, the excitation energy of
this singlet exciton defines the $S=0$ gap. 

\begin{figure}[t]
 \includegraphics[width=0.9\columnwidth,angle=0]{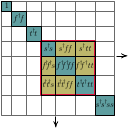}
 \caption{(Color online) General structure of the effective Hamiltonian derived by applying
the generator $\eta_{t:2;f:2;s:1}$, defined in Eq.~\reqn{eq:t2_f2_s1_generator},
 to the initial Hamiltonian~\reqn{eq:IHM_dimer_rep}. 
The ground state, the one-fermion, and the one-triplon sectors are 
decoupled from the remainder of the Hilbert space. 
But the interactions among the one-singlon, the two-fermion, 
 and the two-triplon sector are still present, see red block.}
 \label{fig:eff_hamiltonian}
\end{figure}

Unfortunately, it is not possible to decouple the two-fermion sector from the other sectors
because  the local dimer energies of two fermions, two triplons, and
one singlon are quite close to one another, see Eqs.~\reqn{eq:local_energy_t} and \reqn{eq:local_energy_s}. 
Thus, the $S=0$ gap as well as the low-energy spectrum of the BI phase,
see the next section, are calculated using the generator $\eta^{\phantom{\dagger}}_{t:2;f:2;s:1}$. 
This generator decouples the direct sum of the two-fermion sector, the two-triplon sector, and the
one-singlon sector from other quasiparticle sectors, see Fig.\ \ref{fig:eff_hamiltonian}. 
But the off-diagonal interactions between these three sectors are left out in the generator and thus
they persist in the final effective Hamiltonian. Note that they are nevertheless renormalized
in the course of the CUT.

The structure of the final effective Hamiltonian is schematically shown in Fig.~\ref{fig:eff_hamiltonian}.
The generator $\eta^{\phantom{\dagger}}_{t:2;f:2;s:1}$ can explicitly be written as
\bearr
\eta^{\phantom{\dagger}}_{t:2;f:2;s:1}(\ell) &=& \eta^{\phantom{\dagger}}_{t:0;f:0;s:0}(\ell)
+\eta^{p}_{t:1;f:0;s:0}(\ell) + \eta^{p}_{t:0;f:1;s:0}(\ell) \nn \\
&&\hspace{-1cm}+{\sum_{i+j+k\geq 2}} \left( \hat{\eta}\left[ H_{0,0,1}^{i,j,k}(\ell) \right] 
- {\rm H.c.} \right)X^{i,j,k}_{2,0,0}X^{i,j,k}_{0,2,0} \nn \\
&&\hspace{-1cm}+{\sum_{i+j+k\geq2}} \left( \hat{\eta}\left[ H_{2,0,0}^{i,j,k}(\ell) \right] 
- {\rm H.c.} \right)X^{i,j,k}_{0,2,0} \nn \\
&&\hspace{-1cm}+{\sum_{i+j+k\geq2}} \left( \hat{\eta}\left[ H_{0,2,0}^{i,j,k}(\ell) \right] 
- {\rm H.c.} \right)X^{i,j,k}_{2,0,0},
\label{eq:t2_f2_s1_generator}
\eearr
where the definitions~\reqn{eq:f1_pgenerator} and \reqn{eq:t1_pgenerator} are used in the first 
line. It is seen that the off-diagonal interactions $H_{2,0,0}^{0,0,1}$, $H_{0,2,0}^{0,0,1}$,
and $H_{0,2,0}^{2,0,0}$ are excluded from the generator~\reqn{eq:t2_f2_s1_generator} using
the definition~\reqn{eq:X_def}. 

The $S=0$ gap is calculated by an exact diagonalization in the singlet channel
of the subspace spanned by the states comprising two fermions, two triplons, or one singlon. 
The employed exact diagonalization technique is valid in the thermodynamic limit.
The only restriction required to deal with a finite-dimensional, numerically tractable
problem is the limitation of the distances 
between two quasiparticles, see Refs.~\onlinecite{Fischer10} and \onlinecite{HafezTorbati2014}.
Since this is a two-particle problem we can treat very large relative distances and
find the converged eigenvalues.

\begin{figure}[t]
\includegraphics[width=0.9\columnwidth,angle=-90]{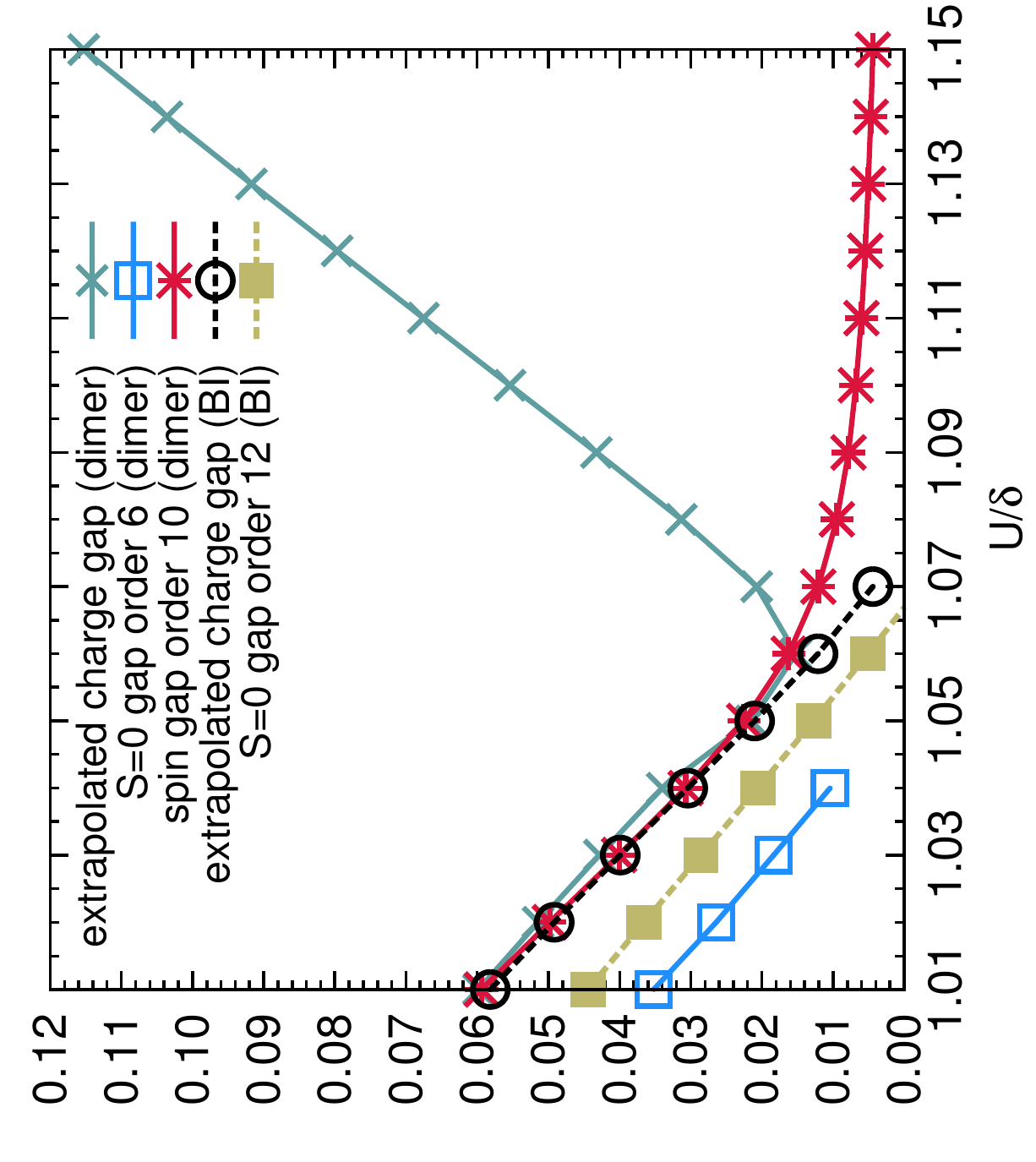}
  \caption{(Color online) The charge gap, the $S=0$ gap, and the spin (triplon) gap of 
the Hamiltonian~\reqn{eq:IHM_res} versus the Hubbard interaction 
$U/\delta$ for $t=0.05\delta$. The results obtained 
from the BI limit (dashed lines up to $U_{c1}\approx 1.067\delta$, value obtained
from the vanishing of the $S=0$ gap) 
and the dimer limit (solid lines) are compared.
}
  \label{fig:gaps}
\end{figure}

Fig.~\ref{fig:gaps} shows various gaps versus the Hubbard interaction $U/\delta$
for $t=0.05\delta$. The results obtained from the BI limit are included for 
comparison; they are valid only up to the first transition point $U_{c1}$ where 
the $S=0$ gap closes. The charge gap~$\Delta_c$ is calculated up to order $12$ in the interdimer 
hopping from the dimerized limit and up to order $20$ in the hopping from the BI limit.
Then the finite order results  
are extrapolated to infinite order by a linear fit to the last four orders as we illustrated
for the GSE in the right panel of Fig.~\ref{fig:GSE}.
From Fig.~\ref{fig:gaps} we see that the charge gap obtained from the dimer limit
and from the BI limit agree well up to $U\approx 1.06\delta$ where the charge gap from the dimer limit
acquires a minimum. We interprete this minimum as an indication 
for the first transition point $U_{c1}$ between the  BI and the SDI phase~\cite{Manmana04,HafezTorbati2014}.

Next, we discuss the $S=0$ gap presented in Fig.~\ref{fig:gaps}. 
In the dimer limit approach, order $6$ is the maximum order that we can reach for this quantity. 
Higher orders are not accessible due to divergence in the flow equations.
This is induced by overlapping different continua which occurs the more often the
more quasiparticles are involved. The same problem occurs for $U>1.04\delta$. The results of the BI 
limit in order $12$ in the hopping parameter are  shown for comparison~\cite{HafezTorbati2014}. Again, 
a divergence of the flow equations prevents us to reach higher orders.

For both limits, the $S=0$ gap 
is smaller than the charge gap indicating an $S=0$ electron-hole bound state, i.e., an exciton,
in the BI phase. The $S=0$ gap obtained from the BI limit at order 12 vanishes at the 
critical interaction $U_{c1}=1.067\delta$ which is very close to the DMRG result of the first transition point $1.065\delta$~\cite{Tincani09}. The $S=0$ 
gap computed from the dimer limit turns out to be too low and in the regime of
interest a divergence of the flow equation occurs which must
be attributed to overlapping continua. 

We recall that for the computation of the $S=0$ bound state 
we have to separate a large subspace made of up to two 
excitations from the remaining Hilbert space, see  Fig.\ \ref{fig:eff_hamiltonian}
and for the generator Eq.\ \ref{eq:t2_f2_s1_generator}. 
This aim appears to be too ambitious. We expect that a more sophisticated calculation of the $S=0$ exciton gap from the dimer limit in finite order will display a non-zero minimum very close to the 
transition point. Only for extrapolated infinite order this minimum will vanish.

Next, we consider the spin gap (triplon gap) which is obtained up to order $10$ 
in the interdimer hopping parameter $\lambda t$
and plotted versus the interaction $U$ in Fig.~\ref{fig:gaps} for $t=0.05\delta$.
The generator used is the one in Eq.\ \ref{eq:t1_generator}.
The results of order $8$ almost coincide with the results of order $10$ especially inside the BI phase. 
We did not perform an extrapolation, as we did for the charge gap,  because 
the triplon gap does not display a clear linear 
behavior versus the inverse order up to order $10$. 
Higher orders would be necessary for an accurate extrapolation to infinite order.

It is seen from  Fig.~\ref{fig:gaps} that the spin gap and the charge gaps are very close up to $U=1.05\delta$. The BI limit analysis shows equal spin and charge gaps up to the transition point 
$U=1.067\delta$ within numerical accuracy. We use this fact to estimate the error in the finite order
calculations of the spin gap. In fact, the spin gap at the transition point $U=1.067\delta$ should be about  $0.007\delta$ smaller to match the results in the BI limit. If we assume that
the spin gap is overestimated by this amount we conclude that the spin mode becomes soft at $U_{c2}=1.10\delta$ indicating the second transition from the SDI to the MI. Indeed, 
this rough estimate is in reasonable agreement with the DMRG result  $U_{c2}\simeq1.085\delta$~\cite{Tincani09}.

In finite order, here order $10$, the spin gap remains finite even for large values
of the Hubbard interaction. This seems to contradicts the fact that a second transition to the
MI phase occurs at larger interaction~\cite{Tincani09,Manmana04}. 
But it must be recalled that the gapless MI phase is unstable versus dimerization, 
that means, dimerization is a relevant perturbation \cite{cross79}. 
Any finite dimerization introduces a finite spin gap in the system. 
By construction, the deepCUT based on the dimer limit introduces
dimerization breaking the symmetry between adjacent bonds, i.e.,
the reflection symmetry about each site. This broken symmetry
is never fully restored in any finite order calculations. Consequently,
a finite spin gap remains. 

In the next section, we derive a low-energy effective Hamiltonian solely in terms of triplon operators 
for larger values of the Hubbard interaction, $U\geq 1.15\delta$. 
This low-energy Hamiltonian is analyzed using a second application of the deepCUT.
In this way, we are able to calculate the spin gap up to much higher orders 
than $10$. The extrapolation of the high order results to infinite
order clearly show the expected tendency towards zero spin gap in the MI phase.

\section{Dispersions}
\label{sec:spectrum}

In this section we investigate the momentum dependent
low-energy excitation spectrum of the restricted IHM~\reqn{eq:IHM_res}
in the BI, in the SDI, and in the MI phase. 
The momentum dependent excitation spectrum of the IHM in the BI phase has been discussed 
in Refs.~\onlinecite{Hafez10b,Hafez11,HafezTorbati2014} from the BI limit.
Here we will corroborate these findings by results obtained based on the dimer limit.
In the SDI and MI phases, the dispersions have not yet been
analyzed quantitatively.

\subsection{Band Insulator Phase}

In the BI phase, the dressed electrons and holes are the elementary excitations of the system. 
These fermionic  quasiparticles with spin $S=1/2$ can form singlet or triplet bound states. 
By starting from the dimer limit, however, we have introduced three different kinds of quasiparticles 
in the system: Fermions, triplons, and singlons. The latter two are of bosonic character.
 It is very interesting to see how well the deepCUT calculations based on the dimer limit
 reproduce the dispersions in the BI phase.

The dispersion of the restricted IHM~\reqn{eq:IHM_res} in the BI phase is obtained by using 
the generator~\reqn{eq:t2_f2_s1_generator} in the 
deepCUT. As discussed in the previous section for the $S=0$ exciton gap, the generator 
$\eta^{\phantom{\dagger}}_{t:2;f:2;s:1}$ maps the initial Hamiltonian~\reqn{eq:IHM_dimer} to an effective Hamiltonian with the general structure shown in Fig.~\ref{fig:eff_hamiltonian}. 
In this effective Hamiltonian, the one-singlon, two-fermion, and two-triplon sectors 
are separated as a whole from other sectors. But there are still off-diagonal 
interactions linking these three sectors among one another.

Because the one-fermion and the one-triplon sectors are decoupled 
from the rest, the fermion and the 
triplon dispersions can be obtained by a simple Fourier transformation. The eigenvalues of the 
Hilbert space composed of the direct sum of the one-singlon, the two-fermion, 
and the two-triplon subspace are calculated by constructing the Hamiltonian matrix for each specific total momentum, 
total charge, total spin, and total magnetic number and performing an ED~\cite{Fischer10,HafezTorbati2014}. 
The employed ED is valid in the thermodynamic limit and we only need 
to restrict the distances between the quasiparticles~\cite{Fischer10,HafezTorbati2014}. 
We focus on the sector with no net total charge where the two electrons are of
different types. The dimension of the Hamiltonian matrix in the $S=0$ channel is $3d+1$ and in the 
$S=1$ channel it is $3d$ where $d$ is the maximum distance between quasiparticles. 
This linear dependence allows us to easily reach very large distances and to find accurate
results for the eigenvalues.

\begin{figure}[t]
\includegraphics[width=0.8\columnwidth,angle=-90]{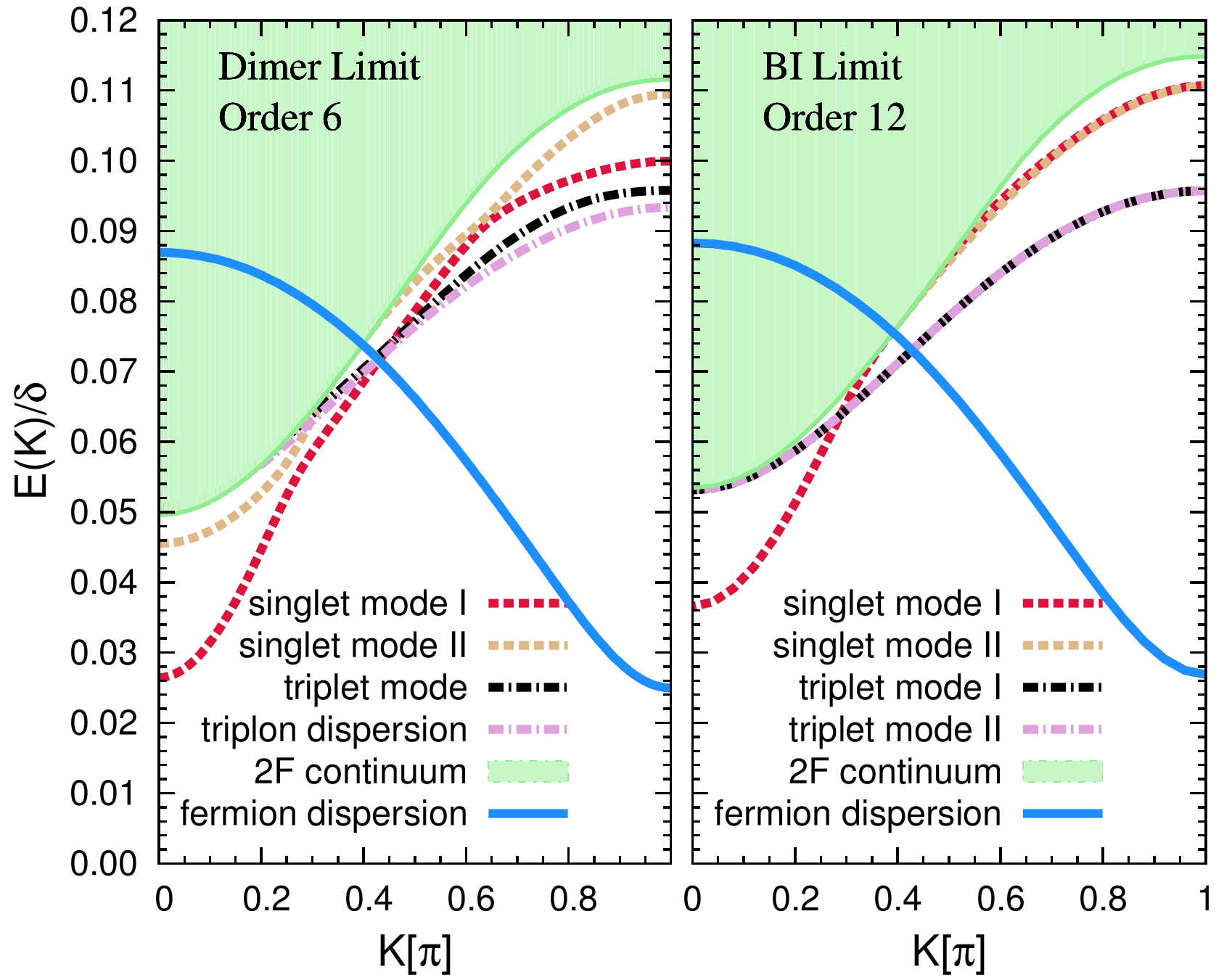}
  \caption{(Color online) The dispersions of Hamiltonian~\reqn{eq:IHM_res} in the BI phase
(hopping parameter $t=0.05\delta$, Hubbard interaction $U=1.02\delta$). 
 The shaded (colored) area indicates the range of the two-fermion (2F) continuum. 
 The left panel shows the results calculated in the dimer limit analysis in order $6$. 
 For comparison, the results from the BI limit in order $12$ are plotted in the right panel. 
 Two singlet bound states are found in both the dimer limit and the BI limit
 calculation. The two triplet modes I and II of the right panel are reproduced in the left panel 
 as a $S=1$ exciton, called triplet mode,  and as a dispersive triplon.}
  \label{fig:BI_disps}
\end{figure}

The dispersions of the restricted IHM~\reqn{eq:IHM_res} for 
the parameters $U=1.02\delta$ and $t=0.05\delta$ are plotted versus the 
total momentum $K$ in Fig.~\ref{fig:BI_disps}.
In this figure, the lattice spacing between the centers of two dimers, which is twice the distance between two sites, is considered as unit of length.
The left panel of Fig.~\ref{fig:BI_disps} shows the results obtained 
from the dimer limit. The BI limit results, which are expected to be more accurate in the BI phase, 
are depicted in the right panel  for comparison.

The BI limit analysis of the IHM is described in Ref.~\onlinecite{HafezTorbati2014} in detail. 
In Ref.~\onlinecite{HafezTorbati2014} the distance between
two {\it sites} is considered as the unit of length. Thus
these results have to be folded to the reduced BZ to compare them with the dimer limit results. 
The fermion dispersion of the BI limit has also to be shifted by $\pi/2$ on the 
momentum axis due to a local transformation applied to the fermion operators, 
see Eq.~({\color{blue} 16}) of Ref.~\onlinecite{HafezTorbati2014}.

Orders $6$ and $12$ are the maximum orders reached
in the dimer limit and  in the BI limit, respectively. 
Both analyses involve the same lattice extension because the range of processes 
taken into account in the deepCUT is proportional to the order of calculations
with a factor of 2 for the dimer limit because the lattice distance between
two dimers is two lattice spacings.

In the right panel of Fig.~\ref{fig:BI_disps}, there appear two singlet
and two triplet bound states in the excitation spectrum obtained from the BI limit. 
The singlet mode II exists 
in the momentum range $\pi/2 \lesssim K\leq \pi$ and its energy coincides with the singlet mode I. 
The two triplet bound states are on top of each other and exist almost in the whole Brillouin
zone. 
The dimer limit also yields two singlet bound states  I and II shown in the left panel 
of Fig.~\ref{fig:BI_disps}. The two singlet modes are not degenerate as
in the BI limit because the dimer approach breaks an additional symmetry
which is not restored completely due to the truncation of the flow equations. 
But in view of this approximation the qualitative agreement of
the results from both limits is satisfactory.
The two degenerate triplet bound states (triplet modes) I and II in the right panel 
appear also in the left panel with an almost quantitative degeneracy.
The agreement of the $S=1$ results from both limits is very good.

\subsection{Spontaneously Dimerized Phase}

The excitation spectrum in the BI phase can be understood well
in terms of electrons and holes and their binding phenomena~\cite{Hafez10b,HafezTorbati2014}.
In the MI phase, the charge degree of freedom appear only at high energies (they are frozen 
at low energies) and the magnetic low-energy excitations
are spinons~\cite{fadde81,mulle81,Karbach97} in the uniform case or triplons
for any dimerization~\cite{cross79,Schmidt03}.
The competition between charge and spin degrees of freedom in the 1D IHM leads 
to an intermediate spontaneously dimerized, insulating phase. 
In this phase, both charge and spin excitations contribute to the 
low-energy spectrum of the system making it difficult to determine quantitatively. 
By construction,  the approach based on the dimer limit 
is especially suited to investigate  the SDI phase of the IHM. 

\begin{figure}[t]
\centering
\includegraphics[width=0.79\columnwidth,angle=-90]{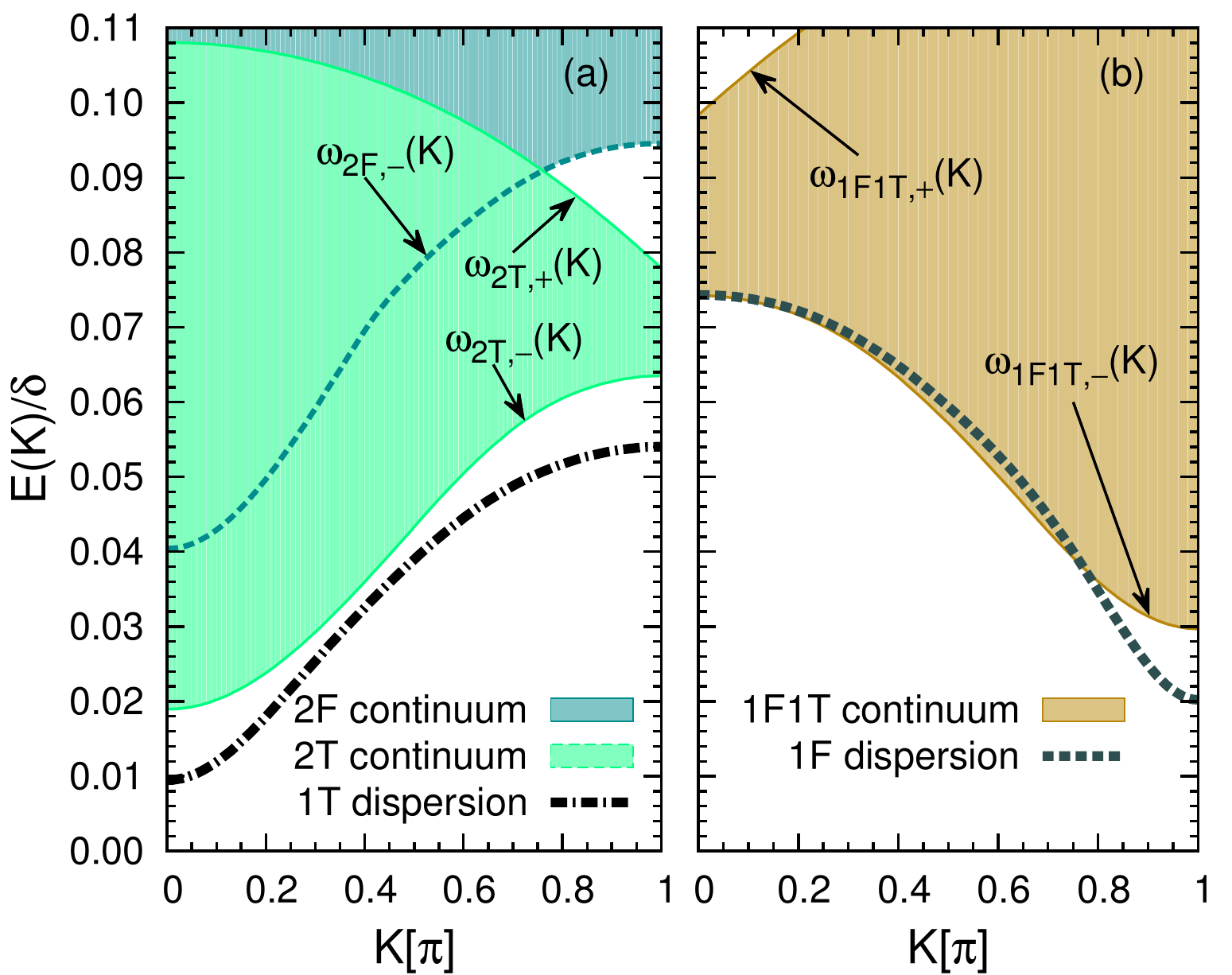}
  \caption{(Color online) The low-energy spectrum of the Hamiltonian~\reqn{eq:IHM_res} for 
  $t=0.05\delta$ and  $U=1.08\delta$ in the SDI phase. The order of the calculation for 
  the fermion (1F) dispersion is $12$ and for the triplon (1T) dispersion it is $10$. 
  The two-fermion (2F) continuum, the two-triplon (2T) continuum, and the fermion-triplon (1F1T)
  continuum are shown as shaded (colored) region. The lower band edge $\omega_{2T,-}(K)$ and the upper 
  band edge $\omega_{2T,+}(K)$ of the two-triplon continuum and the lower band edge $\omega_{2F,-}(K)$ 
  of the two-fermion continuum are shown as well as the lower band edge $\omega_{1F1T,-}(K)$ and the upper 
  band edge $\omega_{1F1T,+}(K)$ of the fermion-triplon continuum. The left panel depicts
  the excitation spectrum with an even number of fermions and the right panel depicts the 
  excitation spectrum with an odd number of fermions.} 
  \label{fig:SDI_disps2}
\end{figure}

We present the results obtained for the fermion dispersion 
$\omega_f(K)$ (1F in Fig.\ \ref{fig:SDI_disps2}) and the triplon dispersion $\omega_t(K)$ 
(1T in Fig.\ \ref{fig:SDI_disps2}) in the SDI phase of the IHM. The 
fermion dispersion is obtained by the generator~\reqn{eq:f1_generator} targeting
the ground state and the one-fermion sector. Similarly, the triplon dispersion is 
calculated by the generator~\reqn{eq:t1_generator} targeting the ground state and 
the one-triplon sector. The order of the calculations is $12$ for the fermion dispersion and
$10$ for the triplon dispersion in the relative interdimer hopping $\lambda$. 
The two-triplon (2T) continuum, the two-fermion (2F) continuum, and the fermion-triplon (1F1T) 
continuum are also depicted in Fig.~\ref{fig:SDI_disps2}.
The hopping parameter and the Hubbard interaction are fixed 
to $t=0.05\delta$ and $U=1.08\delta$. The DMRG data \cite{Tincani09} 
indicate that for $t=0.05\delta$ the SDI phase exists between $U_{c1}=1.065\delta$ and 
$U_{c2}\approx 1.085\delta$ so that for $U=1.080\delta$ we expect the IHM to be in the SDI phase.

The excitation spectrum containing an even number of fermions is plotted in the left panel of 
Fig.~\ref{fig:SDI_disps2}. The right panel of Fig.~\ref{fig:SDI_disps2} indicates the energy 
spectrum with an odd number of fermions.
The dispersions in Fig.~\ref{fig:SDI_disps2} clearly show that \emph{both} 
the spin and the charge excitations contribute 
to the low-lying excitation spectrum of the IHM in the SDI. Concomitantly, the two-particle
continua play an important role. Thus, we indicate the boundaries
of these continua in Fig.~\ref{fig:SDI_disps2} as well. 
The upper band edge of the
two-triplon continuum is denoted by $\omega_{2T,+}(K)$ and the lower one by
$\omega_{2T,-}(K)$. The lower band edge of the
two-fermion continuum is denoted by $\omega_{2F,-}(K)$ which strongly overlaps with the 
two-triplon continuum. The upper fermionic continuum edge lies too
high in energy so that it does not appear in Fig.~\ref{fig:SDI_disps2}.
We have also shown the lower band edge $\omega_{1F1T,-}(K)$ and the upper band edge $\omega_{1F1T,+}(K)$ 
of the fermion-triplon continuum.

The triplon dispersion is maximum at $K=\pi$ and lies energetically always
lower than the two-triplon continuum and the two-fermion continuum so that no decay occurs.
The fermion dispersion takes its minimum at momentum $K=\pi$ lying below the 
fermion-triplon continuum. The fermion dispersion almost coincides with the lower band 
edge of the fermion-triplon continuum for the total momenta $K<0.8\pi$. This will
induce singularities in the spectral densities at the lower band edge of the 
fermion-triplon continuum.

The contributions of the charge excitations to the low-energy spectrum in Fig.~\ref{fig:SDI_disps2} 
show that the low-energy physics of the SDI phase is indeed very difficult, if not impossible, 
to describe by a  purely magnetic effective Hamiltonian. 
The approach using elementary fermionic quasiparticles (electrons and holes)
works fine in the BI, but it is not appropriate to explain the energy spectrum of the SDI shown in Fig.~\ref{fig:SDI_disps2}.
In terms of elementary fermionic quasiparticles the triplon is an $S=1$ exciton.
One can see from Fig.~\ref{fig:SDI_disps2} that this exciton mode has a particularly large binding 
energy. It is given by the energy difference between the triplon dispersion and the lower band edge
of the two-fermion continuum $\omega_{2F,-}(K)$. 
This large binding energy is evidence of a significant attractive electron-hole interaction in the $S=1$ channel. 

In addition, there are also parts of the two-triplon continuum in 
Fig.~\ref{fig:SDI_disps2} which lie below the two-fermion continuum. 
This means that even scattering states of two $S=1$ excitons lie below
the scattering states of two elementary fermionic excitations.
All these observations underline the difficulty to describe the SDI phase 
in terms of electrons and holes as elementary excitations of the IHM. Thus
the description of the SDI from the dimer limit appears to be suitable.

\subsection{Mott Insulator Phase}
\label{subsec:MI}

The low-lying magnetic excitation spectrum of the gapless MI phase in 1D is  described in terms of 
spin-$1/2$ quasiparticles called spinons \cite{Karbach97}. But starting from a model
with some dimerization the natural candidates for the elementary excitations
are triplons \cite{Schmidt03}. For instance, the dispersions, but
also dynamic structure factors, approach the ones of uniform chains
in the limit of vanishing dimerization. Hence, we proceed with the deepCUT
approach to the MI starting from the dimer limit

Fig.~\ref{fig:gaps} shows that the energy difference between the charge gap and the 
spin gap increases upon increasing Hubbard interaction beyond the first transition 
at $U_{c1}$. Clearly, for large values of $U$ the charge fluctuations are very high in energy 
and the triplon fluctuations determine the low-energy physics of the system. Therefore, 
it is a justified first step to derive an effective magnetic low-energy Hamiltonian 
in terms of triplon operators. For this purpose, the generator $\eta_{f:0;s:0}^{\phantom{\dagger}}(\ell)$ is used  to
separate the sector without any fermions or singlons from the sectors which contain
a finite number of fermions and singlons. One should notice that the sector without singlons 
and fermions still includes triplon fluctuations. The generator $\eta_{f:0;s:0}^{\phantom{\dagger}}(\ell)$
is given by
\be
\eta_{f:0;s:0}^{\phantom{\dagger}}(\ell) 
:=  \sum_{j,k} \left( H_{0,0}^{j,k}(\ell) - H^{0,0}_{j,k}(\ell) \right),
\label{eq:f0_s0_generator}
\ee
where $H_{0,0}^{j,k}(\ell)$ stands for the part of the Hamiltonian which annihilates zero number 
of fermions and singlons and creates $j$ fermions and $k$ singlons.
A possible change in the number of triplons is not considered. The application of the generator~\reqn{eq:f0_s0_generator} to the 
initial Hamiltonian~\reqn{eq:IHM_dimer_rep} yields an effective Hamiltonian whose magnetic low-energy
part is decoupled from the high-energy charge sectors. The low-energy physics of the IHM is determined by this effective Hamiltonian expressed only in terms of triplon operators. We stress
that we do not require that this effective Hamiltonian conserves the number of triplons.

\begin{figure}[t]
\includegraphics[width=0.93\columnwidth,angle=-90]{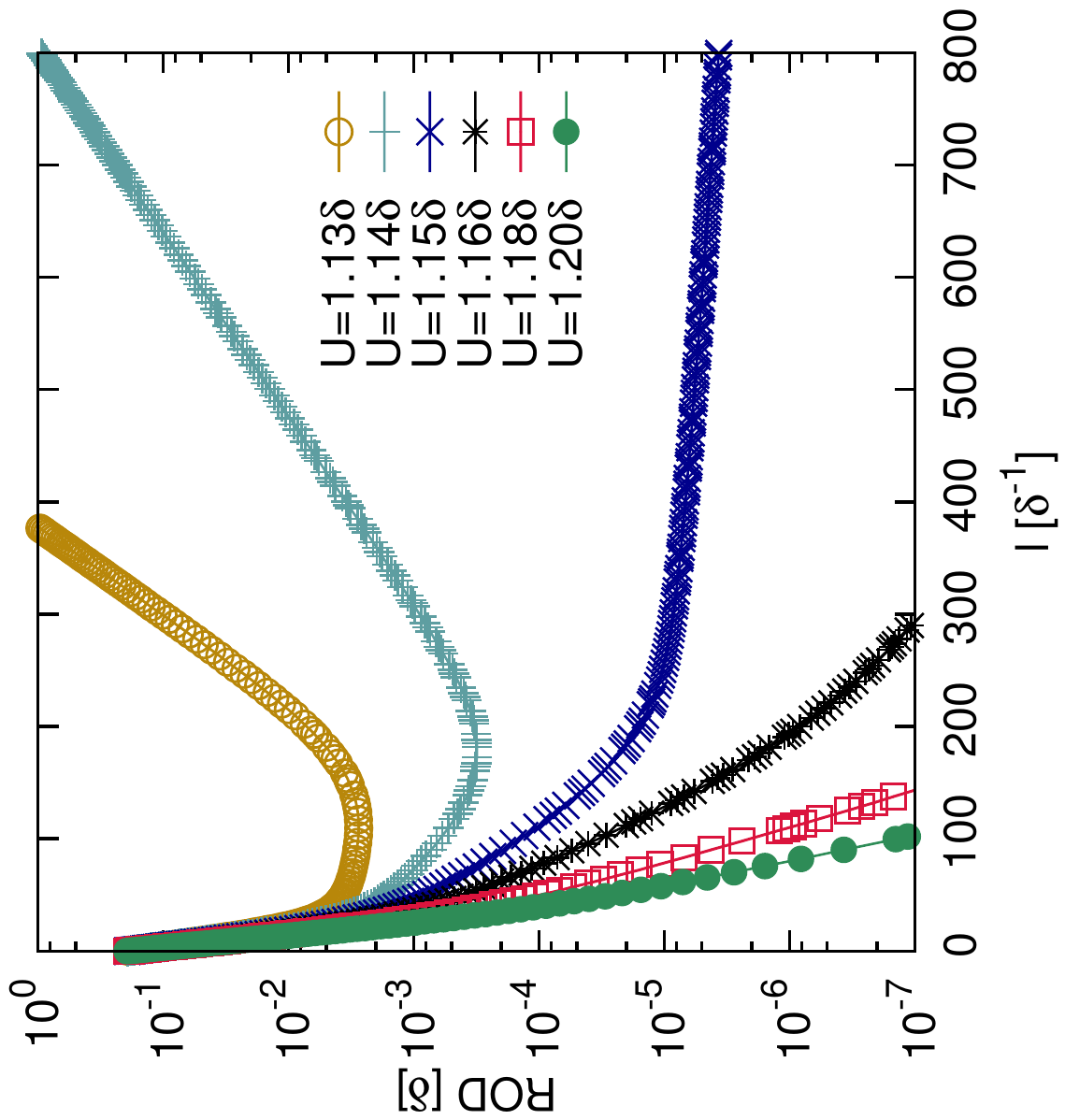}
  \caption{(Color online) The residual off-diagonality (ROD) for the generator $\eta_{f:0;s:0}$ 
as function of the flow parameter $\ell$ for various values of $U$. 
The hopping parameter is fixed to $t=0.05\delta$. The order of the calculations is $10$ targeting all
the monomials composed of triplon operators only.} 
  \label{fig:ROD}
\end{figure}

In order to check the convergence of the flow equations for the generator~\reqn{eq:f0_s0_generator},
the residual off-diagonality (ROD) is plotted in Fig.~\ref{fig:ROD} 
versus the flow parameter $\ell$ for various values of the Hubbard interaction $U$. 
The residual off-diagonality measures the size of the generator: A large ROD
means that the generator is large and vice versa. The rapid vanishing of
the ROD upon increasing $\ell$ signals a good convergence of the CUT.
For further details we refer the reader to  Ref.~\onlinecite{Fischer10}.

The hopping parameter in Fig.~\ref{fig:ROD} is fixed to $0.05\delta$. The order in $\lambda$
of the calculations  is $10$ and we targeted all the monomials composed of triplon operators only.
Fig.~\ref{fig:ROD} shows that the convergence of the flow equations accelerates, i.e., 
improves, for larger Hubbard interactions. We attribute this behavior
 to the larger energy separation between the subspace without
singlon and fermion excitations and the subspace with a finite number of singlons and fermions. 
We expect the effective triplon Hamiltonian to be accurate for $U \geq 1.15\delta$ where a
convergence of ROD is observed. 
The transition from the SDI to the MI phase is predicted
by DMRG to take place at $U_{c2}\approx 1.085\delta$~\cite{Tincani09}. 
Therefore, it appears that the low-energy triplon Hamiltonian has difficulties to describe the system
close to the MI-to-SDI transition. But it should provide a reliable description 
inside the MI phase.

\begin{table}[t]
\caption{\label{tab:eff_ham} All monomials appearing in the effective triplon Hamiltonian 
up to minimal order $2$ in the relative interdimer hopping $\lambda$. 
The index $j$ runs over dimers and the quantum number $m$ takes the 
values $\pm 1$ with $\bar{m}=-m$.}
\begin{ruledtabular}
\begin{tabular}[t]{ | c c c | }
\# & Monomial & Order \\ \hline
$0$	&	$\sum\limits_{j}\mathds{1}$	&	$0$	\\
$1$	&	$\sum\limits_{j} t^\dagger_{j,0} t_{j,0}^{\phantom{\dagger}}$	&	$0$	\\
$2$	&	$\sum\limits_{j,m} t^\dagger_{j,m} t_{j,m}^{\phantom{\dagger}}$	&	$0$	\\
$3$	&	$\sum\limits_{j} \left( t^\dagger_{j,0} t_{j+1,0}^{\phantom{\dagger}} + {\rm H.c.} \right)$	&	$2$	\\
$4$	&	$\sum\limits_{j,m} \left( t^\dagger_{j,m} t_{j+1,m}^{\phantom{\dagger}} + {\rm H.c.} \right)$	&	$2$	\\
$5$	&	$\sum\limits_{j} \left( t^\dagger_{j,0} t_{j+1,0}^{\dagger} + {\rm H.c.} \right)$	&	$2$	\\
$6$	&	$\sum\limits_{j,m} \left( t^\dagger_{j,m} t_{j+1,\overline{m}}^{\dagger} + {\rm H.c.} \right)$	&	$2$	\\
$7$	&	$\sum\limits_{j,m} \left( t_{j,m}^{\phantom{\dagger}} t^\dagger_{j+1,m} t_{j+1,0}^{\phantom{\dagger}}
- t_{j+1,m}^{\phantom{\dagger}} t^\dagger_{j,m} t_{j,0}^{\phantom{\dagger}} 
+ {\rm H.c.} \right)$	&	$2$	\\
$8$	&	$\sum\limits_{j,m} \left( t_{j,0}^{\phantom{\dagger}} t^\dagger_{j+1,m} t_{j+1,m}^{\phantom{\dagger}}
- t_{j+1,0}^{\phantom{\dagger}} t^\dagger_{j,m} t_{j,m}^{\phantom{\dagger}}
+ {\rm H.c.} \right)$	&	$2$	\\
$9$	&	$\sum\limits_{j,m} \left( t_{j,m}^{\phantom{\dagger}} t^\dagger_{j+1,0} t_{j+1,\overline{m}}^{\phantom{\dagger}}
- t_{j+1,m}^{\phantom{\dagger}} t^\dagger_{j,0} t_{j,\overline{m}}^{\phantom{\dagger}}
+ {\rm H.c.} \right)$	&	$2$	\\
$10$	&	$\sum\limits_{j,m} t^\dagger_{j,m} t_{j,m}^{\phantom{\dagger}} t^\dagger_{j+1,m} t_{j+1,m}^{\phantom{\dagger}}$	&	$2$	\\
$11$	&	$\sum\limits_{j} t^\dagger_{j,0} t_{j,0}^{\phantom{\dagger}} t^\dagger_{j+1,0} t_{j+1,0}^{\phantom{\dagger}}$	&	$2$	\\
$12$	&	$\sum\limits_{j,m} t^\dagger_{j,m} t_{j,m}^{\phantom{\dagger}} t^\dagger_{j+1,\overline{m}} t_{j+1,\overline{m}}^{\phantom{\dagger}}$	&	$2$	\\
$13$	&	$\sum\limits_{j,m} \left( t^\dagger_{j,m} t_{j,0}^{\phantom{\dagger}} t^\dagger_{j+1,\overline{m}} t_{j+1,0}^{\phantom{\dagger}} + {\rm H.c.} \right)$	&	$2$	\\
$14$	&	$\sum\limits_{j,m} \left( t^\dagger_{j,m} t_{j,0}^{\phantom{\dagger}} t^\dagger_{j+1,0} t_{j+1,m}^{\phantom{\dagger}}+ {\rm H.c.} \right)$	&	$2$	\\
$15$	&	$\sum\limits_{j,m} \left( t^\dagger_{j,m} t_{j,m}^{\phantom{\dagger}} t^\dagger_{j+1,0} t_{j+1,0}^{\phantom{\dagger}}
+ t^\dagger_{j+1,m} t_{j+1,m}^{\phantom{\dagger}} t^\dagger_{j,0} t_{j,0}^{\phantom{\dagger}} \right)$	&	$2$	\\
\end{tabular}
\end{ruledtabular}
\end{table}

All the monomials in the low-energy triplon Hamiltonian are of 
even order in the perturbative parameter $\lambda t$. Up to order two, they are listed 
in Table~\ref{tab:eff_ham}. 
The lattice extension of each monomial, i.e., the
difference of the index of the rightmost to the leftmost dimer,
 is equal or less than half of its minimal order. 
This means that all monomials with minimal order $2$ can at most act on two adjacent
dimers. This feature helps us to use the same simplification rules as implemented in Ref.~\onlinecite{Krull12} in the subsequent, second  deepCUT applied to 
analyse the effective triplon Hamiltonian.
Henceforth, we switch the formal expansion parameter from $\lambda$ to $\mu = \lambda^2$ 
and present all the results based on orders of $\mu$.

\begin{figure}[t]
\includegraphics[width=0.85\columnwidth,angle=-90]{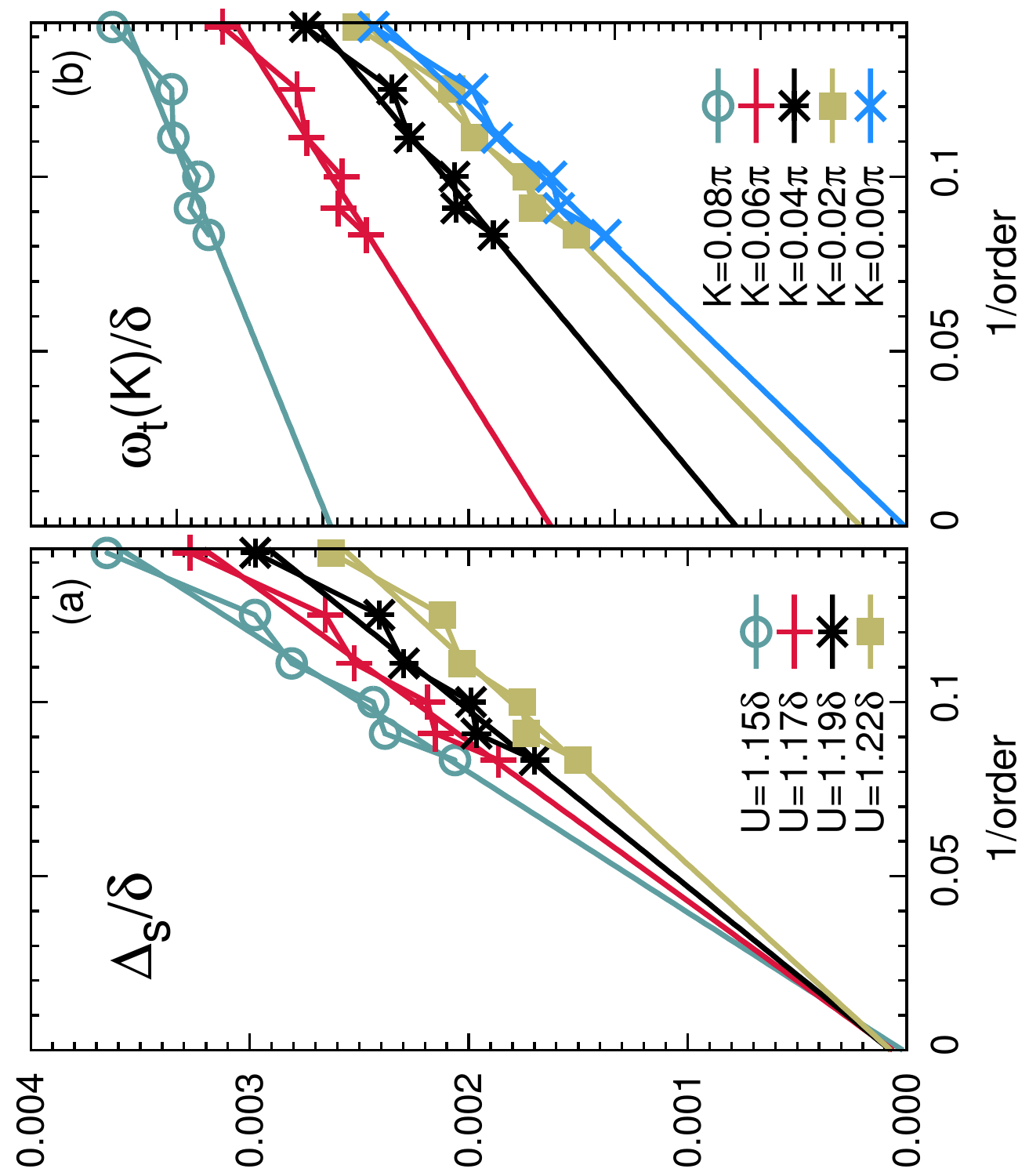}
  \caption{(Color online) Left panel: Spin gap $\Delta_s$ versus the inverse
  order for various $U$. Right panel: 
  The triplon dispersion $\omega_t(K)$ for various fixed total momentum $K$
  versus the inverse order for $U=1.15\delta$. In both
  panels the hopping parameter is set to $t=0.05\delta$. The finite
  order results are extrapolated to infinite order by a linear fit.}
  \label{fig:spin_gap}
\end{figure}

The effective triplon Hamiltonian is mapped by a second application of the deepCUT
to a final effective Hamiltonian whose ground state and one-triplon sector
are separated from higher triplon sectors. This deepCUT allows us to determine the triplon dispersion $\omega_t(K)$.
The minimum of the triplon dispersion occurs at the total momentum $K=0$ so that the
spin (triplon) gap is given by $\Delta_s=\omega_t(0)$. 
Note that one needs to deal with triplon operators only.
This enables us to reach much higher orders compared to the case where all the dimer 
operators~\reqn{eq:dimer_op} matter. Recall that higher orders automatically imply
that processes of longer range are tracked.
We have been able to reach order $12$ in the
expansion parameter $\mu$ equivalent to order $24$ in $\lambda$. 
This maximum order is much higher than 
the order reached in the computation of the spin gap in Fig.~\ref{fig:gaps}.

In the left panel of Fig.~\ref{fig:spin_gap}, the spin gap~$\Delta_s$
is plotted versus the inverse order in $\mu$ for various values of the Hubbard interaction $U$. 
The hopping parameter is fixed to $t=0.05\delta$. 
The finite order results are extrapolated to infinite order by  a linear fit. 
The extrapolated spin gap is lower than $10^{-4}\delta$ indicating the stabilization of the 
gapless MI phase for $U\geq 1.15$. The convergence of the results becomes faster as we increase the 
Hubbard interaction going away from the MI-to-SDI transition point.

We emphasize the importance of the accuracy of the first application of the deepCUT 
in the derivation of the low-energy triplon Hamiltonian. In this first step, we 
target a large number of monomials and a small error may spoil the results obtained
in the second step. For example, by reducing the order of calculations in the first 
step from 10 to 8 we find a slightly negative value for the extrapolated spin gap at $U=1.15$.

\begin{figure}[t]
\includegraphics[width=0.82\columnwidth,angle=-90]{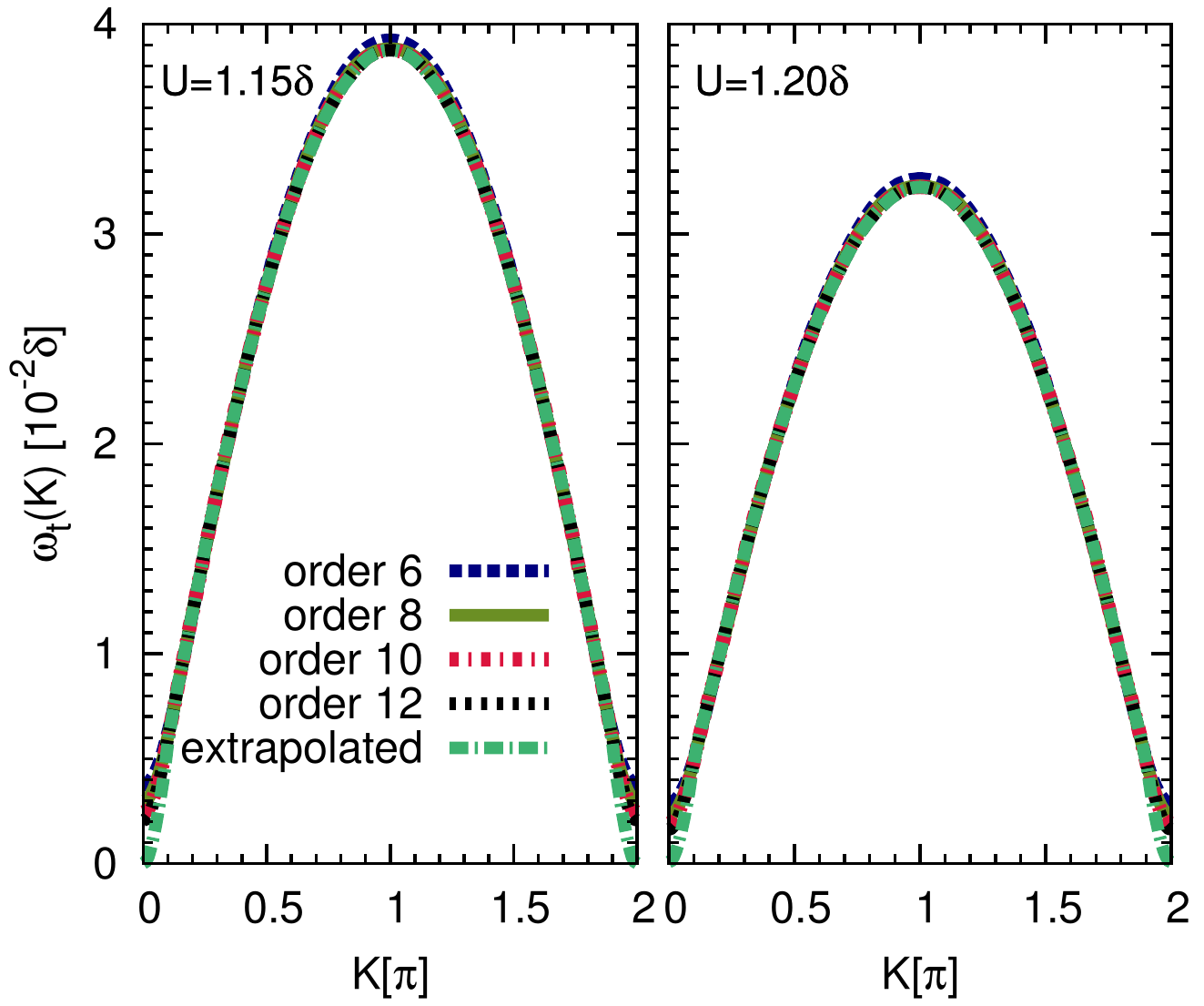}
  \caption{(Color online) Triplon dispersion $\omega_t(K)$ plotted versus
  the total momentum $K$ in the whole Brillouin zone. The hopping parameter
  is  $t=0.05\delta$. In the left panel the Hubbard interaction takes the value $U=1.15\delta$ 
  and in the right panel it is $U=1.20\delta$. The largest deviations between various
  orders occur at $K=0$ and at $K=2\pi$. A numerically gapless triplon dispersion is obtained 
  by extrapolating the finite order results to infinite order by a linear fit in the
  inverse order, see right panel in Fig.\ \ref{fig:spin_gap}.}
  \label{fig:triplon_disp}
\end{figure}

The triplon dispersion $\omega_t(K)$ for various fixed values of the total momentum $K$ 
is plotted versus the inverse order in the right panel of Fig.~\ref{fig:spin_gap}. 
The total momenta are chosen close to $K=0$ 
where the largest deviation between the results of different orders occurs.
The Hubbard interaction $U$ is fixed to $1.15\delta$ and the hopping parameter $t$ is $0.05\delta$. 
Again a linear fit is performed to extrapolate the triplon dispersion to 
infinite order. This extrapolation leads to a numerically gapless triplon dispersion with
a linear behavior $\omega_t(K)\propto |K|$ in the vicinity of $K=0$.

The resulting triplon dispersion is depicted in 
Fig.~\ref{fig:triplon_disp}. The hopping parameter  is again $t=0.05\delta$. 
The Hubbard interaction in the left panel and in the right panel is set to $U=1.15\delta$ 
and $U=1.20\delta$, respectively. In each panel, various finite order results plus the 
extrapolated result are shown for comparison. The triplon dispersions at high 
orders agree well with one another except near the total momenta $K=0$ and $K=2\pi$. 
For these two values of total momentum, the finite order results always lead
to a finite spin gap while the extrapolated result yields a numerically gapless excitation. 

Such a gapless dispersion with linear behavior around $K=0$ and $K=2\pi$ 
is what one expects in a MI phase. It is interesting
to consider the change in the bandwidth of the triplon dispersion in Fig.~\ref{fig:triplon_disp}.
As the Hubbard interaction decreases from $U=1.20\delta$ (right panel) to $U=1.15\delta$ (left panel),
the bandwidth of the triplon dispersion increases. Qualitatively, this finding
can be understood easily by observing that the magnetic
exchange coupling $J$ is $\propto t^2/(U-\delta)$ in  leading order in $t$ in the IHM.
Thus the generic magnetic energy scale decreases upon increasing $U$.

We also analyzed the two-triplon sector of the low-energy triplon Hamiltonian 
searching for possible bound states. It turned out not to be possible to decouple the two-triplon
sector completely from the remaining Hilbert space due to divergencies of the flow equations. 
Thus we proceeded by decoupling the one-triplon sector
and taking the remaining off-diagonal interactions between the two-triplon sector
and the three-triplon sector into account by an exact diagonalization within the Hilbert space made of
up to three triplons~\cite{Fischer10}. 
The calculations in finite order find a weakly bound $S=0$ state. 
But this singlet bound state lies inside the two-triplon continuum as it can 
be constructed from  the \emph{extrapolated} triplon dispersion. 
Thus we conclude that it does not exist as a properly bound state, well-separated
from the two-triplon continuum. Rather we claim that some sort of singular
resonance exists at the lower boundary of the two-triplon continuum.
We presume that it is a divergent power-law as it is found in the sine-Gordon
model at particular values of the interaction \cite{gogol98}.

The vanishing of the spin gap and the stabilization of the MI phase in 
the 1D IHM has been discussed before on the basis of quantum Monte Carlo results~\cite{Wilkens01}
and DMRG results~\cite{Manmana04,Kampf03} without clear-cut conclusion.
Although the position of the second transition point $U_{c2}$ between the SDI and the  MI is not determined accurately in our investigation, our results provide clear evidence that the
MI phase is stable in the \mbox{large-$U$} limit. 
In addition, we obtained quantitative results for the 
magnetic dispersion of the IHM in the MI phase. Our data support the expectation
 that the dimer limit can 
be used as a suitable starting point in CUT-based methods such as deepCUT~\cite{Krull12}, 
perturbative CUT~\cite{Knetter00}, and graph-based CUT~\cite{Yang11} to analyze 
the gapless MI phase in one dimension.

\section{Conclusions}
\label{sec:conclusion}

Strongly correlated systems often give rise to interesting exotic phases
which display unconventional excitations. One strategy to find such phases
is to study systems with a control parameter which switches from one conventional
phase in one limit  to another phase in another limit. In the vicinity of the
phase transition from one conventional phase to the other
the main driving forces counterbalance each other and unexpected mechanisms
may prevail.

In the present work, we study the ionic Hubbard model (IHM) at half filling where a strong
alternation $\delta$ from site to site favors the band insulator (BI) phase while
a strong on-site repulsion $U$ favors the Mott insulator (MI) phase. Thus, the ratio
$g:=U/\delta$ is the control parameter in this case. In one dimension, it is established
that the BI does not become a MI directly upon increasing $g$, but an intermediate spontaneously
dimerized insulating (SDI) phase appears. The goal of the present paper was to
describe the elementary excitations and their dominant interactions
in all these three phases using continuous unitary transformations (CUT). 
This includes the full dependence on the momentum in contrast to many 
purely numerical approaches.
In previous work, this had been achieved in the
BI phase only \cite{Hafez10b,Hafez11,HafezTorbati2014} starting from fermionic quasiparticles.

In order to be able to describe all three phases in one dimension on equal footing,
we chose to start from the dimer limit. This means that we introduce an external
dimerization with a comparably weaker interdimer hopping $\lambda t$ where $0\le \lambda \le 1$.
Thus, the dimensionless parameter $\lambda$ is used to truncate the proliferating number
of terms in the CUTs. The precise scheme how this is done is the deepCUT introduced
earlier \cite{Krull12}. The dimer limit is obviously advantageous in the
description of the spontaneously dimerized phase. But the Mott insulating phase
is also known to be describable from the dimer limit because external dimerization
is a relevant perturbation \cite{Schmidt03}. Last, but not least, 
the fermionic elementary quasiparticles of the BI can also be captured.

Indeed, we could show that the previous results for the fermionic dispersions and
the $S=0$ and $S=1$ excitons in the BI phase are retrieved from the dimer limit.
The dispersions of the fermionic quasiparticles and of the magnetic $S=1$ triplons
agree well with the previous findings. The softening  of the $S=0$ exciton is qualitatively
reproduced, but the quantitative agreement is not as good. This is due to 
the fact that only order 6 could be reached before overlapping continua
spoil the convergence of the flow equations. The critical interaction $U_{c1}$
where the BI phase switches to the SDI phase could be determined nicely
by the intersection of the ground state energies of a deepCUT starting
from the BI limit and from the dimer limit.

In the SDI phase, we could also analyze the dispersions of charge and spin
excitations. We found that it is particularly challenging to sort out these
excitations in the SDI phase because they all have similar energies. Thus
charge and spin degrees of freedom are closely intertwined. This renders
a quantitative description difficult, in particular upon approaching the
transitions at $U_{c1}$ to the BI phase and at $U_{c2}$ to the MI phase.
The value of $U_{c2}$ can only be estimated roughly because the finite spin gap,
induced by the spontaneous dimerization in the SDI phase, vanishes only
very weakly, i.e., exponentially, upon $U\to U_{c2}$ since this transition
is of Kosterlitz-Thouless type \cite{Fabrizio99}.

For the magnetic excitations in the MI, we used the deepCUT in two steps.
First, we systematically derived an effective Hamiltonian in terms of
triplons, i.e., magnetic $S=1$ quasiparticles. In this step, the charge degrees
of freedom are disentangled from the magnetic ones.
In a subsequent, second deepCUT the magnetic effective Hamiltonian is unitarily
transformed such that the number of triplons becomes a conserved quantity.
This allows one to read off the magnetic dispersion directly, for example 
its minimum defining the spin gap.
The extrapolation of the spin gap to infinite order in $\mu=\lambda^2$
reveals that the MI is a phase with massless magnetic excitations.

We summarize that our study yields results for the full
momentum dependence of the charge and spin excitations in all three
phases in one dimension. These results are obtained on equal footing
by introducing an auxiliary dimerization which is sent to zero finally.
The approach is based on a real space representation of the deepCUT.
Since such an approach only captures processes up to a certain range, here
up to 24 lattice spacings, the immediate vicinities of the phase transitions
cannot be described quantitatively. Future work is called for to improve
on this point.

We emphasize that the approach based on the concept of the deepCUT 
has the advantage to be generalizable to higher dimensions. We recall
that the nature of the intermediate phase in two dimensions is
still highly controversial~\cite{Garg06,Kancharla07,Paris07,Craco08,Chen10}.
Surely, the band insulator limit can be used to look for the nature of the modes which
become soft upon increasing interaction, indicating the instability of the BI.
In the Mott insulating phase, the starting point of a long-range ordered
magnet suggests itself. A dimerized limit is conceivable, but so far
no evidence is known to us that this is an ordering pattern likely
to form the intermediate phase in more than one dimension.

\section*{Acknowledgment}
We would like to thank Kai P. Schmidt for fruitful discussions. 
We gratefully acknowledge financial support by the NRW-Forschungsschule
``Forschung mit Synchrotronstrahlung in den Nano- und Biowissenschaften'',
the Mercator Research Center Ruhr ``Elementary excitations and their non-equilibrium 
dynamics in novel materials: From Mott insulators to unconventional superconductors'', 
and the Helmholtz Virtual-Institute ``New states of matter and their excitations''.

\appendix

\section{Simplification Rules}
\label{sec:app}

In this section, we discuss the simplification rules (SRs) that we employed in the dimer 
limit analysis of the IHM~\reqn{eq:IHM_dimer_rep}. 
When we are interested only in some coefficients of the effective Hamiltonian, 
only a small part of the other coefficients is relevant as intermediate result.
Mathematically, the relevance of a monomial can be characterized by its \emph{maximal order} \cite{Krull12}.
The SRs allow us to estimate this maximal order and to discard irrelevant monomials and 
contributions early in the calculation, reducing both runtime and memory consumption drastically.
The concept of SRs and the precise definition of the maximal order are introduced in Ref.~\onlinecite{Krull12}
and the reader is referred to this reference for details.

We distinguish \emph{a priori} and \emph{a posteriori} SRs:
The \emph{a posteriori} SRs are applied to the individual monomials that 
occur by evaluating the commutator of the flow equation \reqn{eq:flow}, 
while the \emph{a priori} SRs are used to estimate the maximal order based 
directly on the arguments that enter the commutator without evaluating it explicitly.
In general, the \emph{a posteriori} SRs eliminate superfluous contributions more thoroughly than their \emph{a priori} counterparts, but the \emph{a priori} SRs can prevent the cumbersome evaluation of the commutator at all when none of the resulting monomials are relevant.
So for the best computational performance, the combined application of both versions of SRs is preferred.

We classify the dimer excitation operators~\reqn{eq:dimer_op} into a boson and a fermion
group. The boson group contains the singlon~\reqn{eq:dimer_op_sing} and the three triplon 
operators~\reqn{eq:dimer_op_trip}. The fermion group includes the four 
fermion operators~\reqn{eq:dimer_op_fer} which act on the left site $f_l^{(\dagger)}$ and on the right site $f_r^{(\dagger)}$ of a dimer with two possible spin quantum numbers.
According to the internal site that a fermion operator acts we define two 
types of fermion operators: `left' and `right'.
The off-diagonal elements in the Hamiltonian~\reqn{eq:IHM_dimer_rep} describe 
various kinds of annihilation and creation processes between and among fermions and bosons.
This makes it difficult to find efficient and flexible SRs especially
if sectors with finite numbers of fermions and/or bosons are targeted. 

In the following, two kinds of SRs are introduced. The first one are the basic SRs
which are not very efficient, but flexible. 
Both, an \emph{a posteriori} and an \emph{a priori} version exist.
They can be applied if sectors with a specific number 
of bosons and fermions are targeted. In the basic SRs, only the number of creation and 
annihilation operators of each monomial is considered. 

In the second kind of SRs, called extended SRs,
the lattice structure of the monomials is also taken into account. 
Here, we derive an \emph{a posteriori} version only.
The extended
SR works efficiently for the ground state and to some extent for the one-fermion 
sector. It is also used for the derivation of the low-energy triplon Hamiltonian where
all the triplon operators are targeted. But the present extended SR needs to be 
generalized if higher fermion sectors or sectors with a mixed number of bosons and 
fermions are targeted.

Before describing the basic and the extended SRs, let us consider the 
general structure of off-diagonal terms in the Hamiltonian~\reqn{eq:IHM_dimer_rep}. 
We emphasize that the first order generator terms are sufficient to consider for deriving
the SRs although higher order terms with more complex structures also appear during 
the flow~\cite{Krull12}. On the one hand, the more complex structure allows
for more complicated cancellations. On the other hand, the higher order limits
the possible effect for given targeted order. In other words, the
more complex term can be understood as being built by iterated 
multiplication or commutation with the first order generator term.

Among  the first order generator terms, only 
the three structures following are important to determine the maximal orders
\begin{subequations}
\label{eq:app:gen}
\begin{align}
\label{eq:app:gen1}
\eta^{(a)} &\propto \sum_j f^{\dagger}_{j;l} f^{\dagger}_{j+1;r} + {\rm H.c.}, 
\allowdisplaybreaks[4] \\
\label{eq:app:gen2}
\eta^{(b)} &\propto \sum_j b^{\dagger}_{j} f^{\phantom{\dagger}}_{j;p} 
f^{\dagger}_{j+1;p}+ {\rm H.c.} \quad;\quad p=l,r, 
\allowdisplaybreaks[4] \\
\label{eq:app:gen3}
\eta^{(c)} &\propto \sum_j b^{\dagger}_{j} b^{\dagger}_{j+1} f^{\phantom{\dagger}}_{j;l} f^{\phantom{\dagger}}_{j+1;r} + {\rm H.c.},
\end{align}
\end{subequations}
where the boson operator $b^{\dagger}$ stands either for a singlon or for a triplon operator and we omitted the spin index of the fermion operators to lighten the notation
and because they play no role in the following considerations.

\subsection{The Basic Simplification Rules}

We start our derivation with the basic \emph{a posteriori} simplification rule.
In the first step, we focus on the number of annihilation and creation operators 
which can be canceled by commutation with the generator structures~\reqn{eq:app:gen}. The 
first term $\eta^{(a)}$ can cancel two fermion creation or annihilation operators only if 
they act on {\it different} intradimer positions which are 
either the left $p=l$ or the right site $p=r$. 
In other words, the generator $\eta^{(a)}$ can only cancel two fermion operators 
of different types.
This reflects the conservation of the total charge of the system.
Two creation or annihilation operators of the same type need
at least two commutations with the generator $\eta^{(a)}$ to be canceled. 
This property allows us to make the SRs dependent on the fermion type
using the same strategy as in Ref.~\onlinecite{HafezTorbati2014}. 
The net effect of the second term $\eta^{(b)}$ is to cancel
one boson operator. The third term $\eta^{(c)}$ transforms two boson operators into two 
fermion operators of different types.

We aim at finding an upper bound for the maximal order of a monomial $A$ if  sectors 
with up to $q_b$ bosons {\it and} up to $q_f$ fermions are targeted up to order $n$. 
We focus on the creation operators supposing
 that the monomial $A$ creates $c_b$ boson and $c_f$ fermions.
The annihilation operators can be treated in the same way.
At first, we discuss the situation where no fermions are targeted,
but $q_b$ bosons. 
Then we aim at keeping $q_f$ fermion operators of the monomial $A$ such that the 
maximal order is overestimated. 
The number of boson operators which have to be canceled 
reads as
\be
c'_b :=  {\rm max}(c_b-q_b,0).
\label{eq:app:cb}
\ee
If $c'_b$ is even,  we transform all these boson operators to fermion operators using $\eta^{(c)}$.
If $c'_b$ is odd, the even number $c'_b-1$ of boson operators are transformed into fermion operators 
and the remaining boson operator is canceled by $\eta^{(b)}$. This procedure requires $\left\lceil \frac{c'_b}{2} \right\rceil$ commutations and produces $2\left\lfloor \frac{c'_b}{2} \right\rfloor$ additional fermion operators. We use the ceiling brackets $\lceil\cdot\rceil$ for the smallest integer
larger or equal to the argument while the flooring brackets $\lfloor\cdot\rfloor$ stand for
the largest integer smaller or equal to the argument in the brackets.
Half of these additional fermion operators are of the type $f^{\phantom{\dagger}}_{l}$ and the other half are of type $f^{\phantom{\dagger}}_{r}$. Therefore, the total number of fermion operators of 
type `left' and `right' which have to be canceled is given by
\begin{subequations}
\label{eq:app:cf}
 \begin{align}
  c'_{f_l} &= c_{f_l} + \left\lfloor \frac{c'_b}{2} \right\rfloor, \allowdisplaybreaks[4] \\
  c'_{f_r} &= c_{f_r} +  \left\lfloor \frac{c'_b}{2} \right\rfloor,
 \end{align}
\end{subequations}
where $c_{f_l}$ and $c_{f_r}$ are the initial numbers of fermion operators of type `left' and 
`right' in the monomial~$A$. 

The fermion operators can be canceled by commutations 
with the term~$\eta^{(a)}$.
This term always cancels two fermions of different types, i.e., on different intradimer positions. 
Fermion operators of the same type need one commutation each. 
Hence, the number of commutations necessary to eliminate all the boson~\reqn{eq:app:cb}
and the fermion~\reqn{eq:app:cf} operators is given by
\be
K_{0,q_b}^{c} = \left\lceil \frac{c'_b}{2} \right\rceil  + {\rm max}(c'_{f_l}, c'_{f_r}).
\ee
This equation can be generalized to $K_{q_f,q_b}^{c}$ if sectors with up to $q_b$ bosons and 
$q_f$ fermions are targeted. In this case, one needs to keep $q_f$ fermion operators of 
monomial $A$ such that the number of commutations $K_{q_f,q_b}^{c}$ is minimized.

We divide the fermion operators into pairs. Each pair contains one `left' and one `right' operator.
First, we keep the fermion operators which do not form pairs.
In this way, one can save one commutation for each fermion operator. 
The remaining fermion operators are all in pairs. This saves one commutation for 
each pair of operators. In this way, we obtain 
\be
K_{q_f,q_b}^{c} = K_{0,q_b}^{c} - d_1^c - \left\lceil \frac{d_2^c}{2} \right\rceil,
\ee
where $d_1^c$ and $d_2^c$ are defined as
\begin{subequations}
\begin{align}
 d_1^c & := {\rm min}\left(q_f, \left| c'_{f_l}-c'_{f_r} \right| \right), 
 \allowdisplaybreaks[4] 
 \\
 d_2^c & := {\rm min}\left(q_f-d_1^c, c'_{f_l}+c'_{f_r}-d_1^c \right).
\end{align}
\end{subequations}
The annihilation part of monomial $A$ can be analyzed in the same way and leads to $K_{q_f,q_b}^{a}$.
Finally, the upper bound for the maximal order of the monomial $A$ is given by~\cite{Krull12}
\be
\widetilde{O}_{\rm max}(A) = n- K_{q_f,q_b}^{c} - K_{q_f,q_b}^{a},
\label{eq:app:omax_basic}
\ee
where $n$ is the order of calculations. The monomial $A$ has no effect on the targeted quantities
up to order $n$ and can be neglected if 
\be
\widetilde{O}_{\rm max}(A) < O_{\rm min}(A).
\label{eq:app:basic_aposteriori}
\ee
We refer to this analysis as the basic \emph{a posteriori} SR.

Now, we
explain the basic \emph{a priori} SR for the commutator $\left[ T,D \right]=TD-DT$. 
We focus on the product $TD$. The product $DT$ can be treated in the same way. 
All we need to do is to bound the number of creation and annihilation operators 
which remain after normal-ordering from below.

Suppose $c_{T}^{i}$ and $a_{T}^{i}$ are the numbers 
of creation and annihilation operators of type~$i$, respectively, in the monomial $T$. 
The index $i$ refers to the eight possible operators (singlon, triplons, and fermions) 
that can appear on a dimer, cf.\ Tab.\ \ref{tab:dimer}.
Similarly, there are $c_{D}^{i}$ creation 
operators and $a_{D}^{i}$ annihilation operators of type~$i$ in  the monomial $D$.
Only operators of the same type can cancel each other in the process of normal-ordering.
Therefore, the number of creation and annihilation operators
of type $i$ of the product $TD$ can be bounded from below by
\begin{subequations}
\begin{align}
 c_{TD}^{i} &\geq \tilde{c}_{TD}^{i} := c_{T}^{i} + c_{D}^{i} - s_{TD}^{i}, 
 \allowdisplaybreaks[4] \\
 a_{TD}^{i} &\geq \tilde{a}_{TD}^{i} := a_{D}^{i} + a_{T}^{i} - s_{TD}^{i}, 
 \allowdisplaybreaks[4]
\end{align}
\end{subequations}
where $s_{TD}^i := \min\left( a_{T}^{i}, c_{D}^{i} \right)$. Subsequently, for 
the number of boson operators and the number of `left' and `right' fermion operators 
we find 
\begin{subequations}
\label{eq:app:op_estim}
\begin{align}
 c^{b}_{TD} &\geq \tilde{c}^{b}_{TD} := \sum_{i \in {\rm b} } \tilde{c}^{i}_{TD}, 
 \allowdisplaybreaks[4] \\
 c^{f_r}_{TD} &\geq \tilde{c}^{f_r}_{TD} := \sum_{i \in f_r } \tilde{c}^{i}_{TD}, 
 \allowdisplaybreaks[4] \\
 c^{f_l}_{TD} &\geq \tilde{c}^{f_l}_{TD} := \sum_{i \in f_l } \tilde{c}^{i}_{TD}, 
 \allowdisplaybreaks[4]
\end{align}
\end{subequations}
where $b$ stands for bosons and $f_l$ and $f_r$ stand for `left' and `right' fermions. Analogous relations as~\reqn{eq:app:op_estim} are valid for the annihilation operators in $TD$. 

On the basis of the numbers $c^{b}_{TD}$, $c^{f_r}_{TD}$, $c^{f_l}_{TD}$, 
$a^{b}_{TD}$, $a^{f_r}_{TD}$, and $a^{f_l}_{TD}$, we can estimate 
the maximal order of the product $TD$ using the relation~\reqn{eq:app:omax_basic}.
Finally, the commutator $\left[ T,D \right]$ has no effect on the targeted 
quantities up to order $n$ and can be ignored if
\be 
\max \left( \widetilde{O}_{\rm max}(TD), \widetilde{O}_{\rm max}(DT) \right) < 
O_{\rm min} (T) + O_{\rm min} (D).
\ee
This basic \emph{a priori} SR can be used in addition to the basic \emph{a posteriori}
SR \reqn{eq:app:basic_aposteriori} or in addition to the extended \emph{a posteriori} SR 
that is presented in the next subsection, in order to increase the speed of the deepCUT 
algorithm.

\subsection{The Extended A Posteriori Simplification Rule}

The upper bound~\reqn{eq:app:omax_basic} for the maximal order of the monomial $A$ 
can be lowered by taking into account the lattice structure of the generator terms~\reqn{eq:app:gen}.
The first term $\eta^{(a)}$ cancels two fermions of different type,
 but only on nearest-neighbor (\NN) dimers. 
The third term $\eta^{(c)}$ transforms two \NN\ bosons into two \NN\
 fermions of different type.
Special attention has to be paid to the second term $\eta^{(b)}$. This term can cancel one boson
accompanied with a \NN\ hopping process for a fermion. This additional hopping process makes it difficult to derive an efficient extended SR as done in Refs.~\onlinecite{Krull12,HafezTorbati2014}. 
In the following, we derive a lower bound for the number of commutations required to cancel all the fermion and boson creation operators of monomial $A$.
The annihilation part of monomial $A$ can be treated in the same way.

Similar to Ref.~\onlinecite{Krull12}, we consider the monomial $A$ and
split the cluster of sites on which creation operators
act into different {\it linked} subclusters.
We define $K\left[ \mathcal{C} \right]$ as the number of commutations needed to cancel all operators 
of the linked subcluster $\mathcal{C}$. We have to bound this number of commutations from below in 
order to find an upper bound for the maximal order of monomial~$A$. 
The size of each linked subcluster can be 
reduced based on the following inequalities
\begin{subequations}
\label{eq:app:ineq}
 \begin{align}
\label{eq:app:ineqI}
  K\!\left[\boldsymbol{\otimes\!\! -\!\! \oplus\!\!-\!}\mathcal{C}' \right] 
&= K\!\left[ \boldsymbol{\otimes\!\! -\!\! \oplus} \right] + K\!\left[ \mathcal{C}' \right]
\geq 1+ K\!\left[ \mathcal{C}' \right], \\
\label{eq:app:ineqII}
  K\!\left[\boldsymbol{\CIRCLE\!\! -\!\! \CIRCLE\!\!-\!}\mathcal{C}' \right] 
  &= K\!\left[ \boldsymbol{\CIRCLE\!\! -\!\! \CIRCLE} \right] + K\!\left[ \mathcal{C}' \right]
\geq 2+ K\!\left[ \mathcal{C}' \right], \\ 
\label{eq:app:ineqIII}
  K\!\left[\boldsymbol{\CIRCLE\!\! -\!\! \otimes\!\!-\!}\mathcal{C}' \right] 
  &= K\!\left[ \boldsymbol{\CIRCLE\!\! -\!\! \otimes} \right] + K\!\left[ \mathcal{C}' \right]
\geq 2+ K\!\left[ \mathcal{C}' \right], \\ 
\label{eq:app:ineqIV}
  K\!\left[\boldsymbol{\otimes\!\! -\!\! \CIRCLE\!\!-\!}\mathcal{C}' \right] 
  &\geq 1+ K\!\left[ \boldsymbol{\otimes\!\!-\!} \mathcal{C}' \right], \\ 
\label{eq:app:ineqV}
  K\!\left[\boldsymbol{\otimes\!\! -\!\! \otimes\!\!-\!}\mathcal{C}' \right] 
  &\geq 1+ K\!\left[ \boldsymbol{\otimes\!\!-\!} \mathcal{C}' \right],
 \end{align}
\end{subequations}
where the symbols $\boldsymbol{\otimes}$ and $\boldsymbol{\oplus}$ denote the two 
possible fermion operators on a dimer, the symbol $\CIRCLE$ stands for a boson operator 
on a dimer, and $\mathcal{C}'$ stands for the remaining part of the initial subcluster.
For instance, on the left hand side of the first equation~\reqn{eq:app:ineqI}
the two \NN\ fermion operators of different 
type \mbox{$\boldsymbol{\otimes\!\!-\!\! \oplus}$} are linked to the remaining part~$\mathcal{C}'$. 
The validity of Eqs.~\reqn{eq:app:ineq} can be derived on the basis of the structure of the generator terms~\reqn{eq:app:gen}. 

As an example, let us illustrate that one cannot
cancel the linked fermion and boson operators \mbox{$\boldsymbol{\otimes\!\!-\!\!\CIRCLE}$} 
in Eq.~\reqn{eq:app:ineqIV}. Consider a linked cluster of the form 
\be
\mathcal{C} = \boldsymbol{\otimes\!\!-\!\!\underbrace{\CIRCLE\!\!-\!\!\CIRCLE\!\!-
\cdots -\!\!\CIRCLE\!\!-\!\!\CIRCLE}_n\!\!-\!\oplus},
\label{eq:app:examp}
\ee
where two different fermion operators at the two outer ends of the cluster $\mathcal{C}$ are linked 
by $n$ boson operators. By $n$ applications of the generator term~\reqn{eq:app:gen2}, one can 
cancel all the boson operators  and bring the two fermion operators to adjacent dimers. This pair
of \NN\ fermion operators can also be canceled by one additional commutation with the generator term~\reqn{eq:app:gen1}.
Hence, the total number of commutations required to cancel the cluster~\reqn{eq:app:examp} is $n+1$.

The scheme in Eqs.~\reqn{eq:app:ineq} reduces a linked subcluster of operators
to at most an individual fermion or boson operator. 
A single boson operator requires two commutations and a single
fermion operator one commutation to be canceled. In this manner, we can find the 
minimum number of commutations $\widetilde{K}\left[\mathcal{C}\right]$ 
to cancel all operators of the linked subcluster $\mathcal{C}$. Then, the minimum 
number of commutations necessary to cancel all the creation operators of monomial $A$ is given by
\be
K_{0,0}^c = \sum_{\mathcal{C}} \widetilde{K}\!\left[ \mathcal{C} \right],
\ee  
where the sum runs over all linked subcluster of creation operators in the 
monomial $A$. Similarly, we can analyze the annihilation operators of the monomial $A$. 
Finally, the maximal order is calculated by Eq.~\reqn{eq:app:omax_basic}.
We refer to this analysis as extended \emph{a posteriori} SR.
This extended \emph{a posteriori} SR is applied in three different cases that 
we discuss in the following.

\subsubsection{The ground state and the one-fermion sector}
For the ground state, one can make the scheme presented in Eq.~\reqn{eq:app:ineq} spin-dependent. 
This takes into account that the two fermion operators in Eq.~\reqn{eq:app:ineqI} can be 
canceled only if they have {\it different} spin. In this case, we should take care
of the spin of the fermion operator in Eq.~\reqn{eq:app:ineqIV}. The spin of the fermion 
operator will be changed from the left to the right hand side in Eq.~\reqn{eq:app:ineqIV} 
if the boson operator is a triplon with the magnetic number $\pm 1$.

This spin-dependent extended SR saves a factor of about $8$ in the number of representatives compared 
to the basic SR if the ground state is targeted.  We also used this extended SR to describe 
the one-fermion sector based on the simple estimate $K_{1,0}=K_{0,0}-1$ for both the creation and 
the annihilation operators. This relation can always be used in Eq.~\reqn{eq:app:omax_basic} because a 
fermion operator requires at most one commutation to be canceled. 
For higher fermion sectors, however, such a simple estimate 
does not work efficiently and further modifications are required.

\subsubsection{Derivation of the triplon Hamiltonian}
The extended \emph{a posteriori} SR is employed also in the derivation of the effective triplon
Hamiltonian in subsection \ref{subsec:MI}. In this deepCUT application, all 
triplon operators are targeted and we need to cancel the singlon and the fermion operators.
The cluster of creation and annihilation operators are made of singlon and fermion operators
only, ignoring the triplon operators. In this case, the extended SR can not 
be made spin-dependent, as in the ground state case, because two fermions with the same spin can be 
transformed into a triplon operator by one commutation.

The number of representatives which remain
when applying this extended SR differs by a factor of about 2 from
the real number of representatives which we need to describe
the triplon Hamiltonian up to order 10. 
This indicates that the extended SR is working incredibly well also for the derivation
of the effective triplon Hamiltonian.

% \bibliographystyle{apsrev4-1}
% \bibliography{references}

\begin{thebibliography}{47}%
\makeatletter
\providecommand \@ifxundefined [1]{%
 \@ifx{#1\undefined}
}%
\providecommand \@ifnum [1]{%
 \ifnum #1\expandafter \@firstoftwo
 \else \expandafter \@secondoftwo
 \fi
}%
\providecommand \@ifx [1]{%
 \ifx #1\expandafter \@firstoftwo
 \else \expandafter \@secondoftwo
 \fi
}%
\providecommand \natexlab [1]{#1}%
\providecommand \enquote  [1]{``#1''}%
\providecommand \bibnamefont  [1]{#1}%
\providecommand \bibfnamefont [1]{#1}%
\providecommand \citenamefont [1]{#1}%
\providecommand \href@noop [0]{\@secondoftwo}%
\providecommand \href [0]{\begingroup \@sanitize@url \@href}%
\providecommand \@href[1]{\@@startlink{#1}\@@href}%
\providecommand \@@href[1]{\endgroup#1\@@endlink}%
\providecommand \@sanitize@url [0]{\catcode `\\12\catcode `\$12\catcode
  `\&12\catcode `\#12\catcode `\^12\catcode `\_12\catcode `\%12\relax}%
\providecommand \@@startlink[1]{}%
\providecommand \@@endlink[0]{}%
\providecommand \url  [0]{\begingroup\@sanitize@url \@url }%
\providecommand \@url [1]{\endgroup\@href {#1}{\urlprefix }}%
\providecommand \urlprefix  [0]{URL }%
\providecommand \Eprint [0]{\href }%
\providecommand \doibase [0]{http://dx.doi.org/}%
\providecommand \selectlanguage [0]{\@gobble}%
\providecommand \bibinfo  [0]{\@secondoftwo}%
\providecommand \bibfield  [0]{\@secondoftwo}%
\providecommand \translation [1]{[#1]}%
\providecommand \BibitemOpen [0]{}%
\providecommand \bibitemStop [0]{}%
\providecommand \bibitemNoStop [0]{.\EOS\space}%
\providecommand \EOS [0]{\spacefactor3000\relax}%
\providecommand \BibitemShut  [1]{\csname bibitem#1\endcsname}%
\let\auto@bib@innerbib\@empty
%</preamble>
\bibitem [{\citenamefont {Gebhard}(1997)}]{Gebhard97}%
  \BibitemOpen
  \bibfield  {author} {\bibinfo {author} {\bibfnamefont {F.}~\bibnamefont
  {Gebhard}},\ }\href@noop {} {\emph {\bibinfo {title} {The Mott
  Metal-Insulator Transition}}}\ (\bibinfo  {publisher} {Springer},\ \bibinfo
  {address} {Berlin},\ \bibinfo {year} {1997})\BibitemShut {NoStop}%
\bibitem [{\citenamefont {Imada}\ \emph {et~al.}(1998)\citenamefont {Imada},
  \citenamefont {Fujimori},\ and\ \citenamefont {Tokura}}]{Imada98}%
  \BibitemOpen
  \bibfield  {author} {\bibinfo {author} {\bibfnamefont {M.}~\bibnamefont
  {Imada}}, \bibinfo {author} {\bibfnamefont {A.}~\bibnamefont {Fujimori}}, \
  and\ \bibinfo {author} {\bibfnamefont {Y.}~\bibnamefont {Tokura}},\ }\href
  {\doibase 10.1103/RevModPhys.70.1039} {\bibfield  {journal} {\bibinfo
  {journal} {Rev. Mod. Phys.}\ }\textbf {\bibinfo {volume} {70}},\ \bibinfo
  {pages} {1039} (\bibinfo {year} {1998})}\BibitemShut {NoStop}%
\bibitem [{\citenamefont {des Cloizeaux}\ and\ \citenamefont
  {Pearson}(1962)}]{Cloizeaux62}%
  \BibitemOpen
  \bibfield  {author} {\bibinfo {author} {\bibfnamefont {J.}~\bibnamefont {des
  Cloizeaux}}\ and\ \bibinfo {author} {\bibfnamefont {J.~J.}\ \bibnamefont
  {Pearson}},\ }\href {\doibase 10.1103/PhysRev.128.2131} {\bibfield  {journal}
  {\bibinfo  {journal} {Phys. Rev.}\ }\textbf {\bibinfo {volume} {128}},\
  \bibinfo {pages} {2131} (\bibinfo {year} {1962})}\BibitemShut {NoStop}%
\bibitem [{\citenamefont {Faddeev}\ and\ \citenamefont
  {Takhtajan}(1981)}]{fadde81}%
  \BibitemOpen
  \bibfield  {author} {\bibinfo {author} {\bibfnamefont {L.~D.}\ \bibnamefont
  {Faddeev}}\ and\ \bibinfo {author} {\bibfnamefont {L.~A.}\ \bibnamefont
  {Takhtajan}},\ }\href@noop {} {\bibfield  {journal} {\bibinfo  {journal}
  {Phys. Lett.}\ }\textbf {\bibinfo {volume} {85A}},\ \bibinfo {pages} {375}
  (\bibinfo {year} {1981})}\BibitemShut {NoStop}%
\bibitem [{\citenamefont {Cross}\ and\ \citenamefont {Fisher}(1979)}]{cross79}%
  \BibitemOpen
  \bibfield  {author} {\bibinfo {author} {\bibfnamefont {M.~C.}\ \bibnamefont
  {Cross}}\ and\ \bibinfo {author} {\bibfnamefont {D.~S.}\ \bibnamefont
  {Fisher}},\ }\href@noop {} {\bibfield  {journal} {\bibinfo  {journal} {Phys.
  Rev. B}\ }\textbf {\bibinfo {volume} {19}},\ \bibinfo {pages} {402} (\bibinfo
  {year} {1979})}\BibitemShut {NoStop}%
\bibitem [{\citenamefont {Uhrig}\ \emph {et~al.}(1999)\citenamefont {Uhrig},
  \citenamefont {Sch\"onfeld}, \citenamefont {Laukamp},\ and\ \citenamefont
  {Dagotto}}]{uhrig99a}%
  \BibitemOpen
  \bibfield  {author} {\bibinfo {author} {\bibfnamefont {G.~S.}\ \bibnamefont
  {Uhrig}}, \bibinfo {author} {\bibfnamefont {F.}~\bibnamefont {Sch\"onfeld}},
  \bibinfo {author} {\bibfnamefont {M.}~\bibnamefont {Laukamp}}, \ and\
  \bibinfo {author} {\bibfnamefont {E.}~\bibnamefont {Dagotto}},\ }\href@noop
  {} {\bibfield  {journal} {\bibinfo  {journal} {Eur. Phys. J. B}\ }\textbf
  {\bibinfo {volume} {7}},\ \bibinfo {pages} {67} (\bibinfo {year}
  {1999})}\BibitemShut {NoStop}%
\bibitem [{\citenamefont {Knetter}\ and\ \citenamefont
  {Uhrig}(2000)}]{Knetter00}%
  \BibitemOpen
  \bibfield  {author} {\bibinfo {author} {\bibfnamefont {C.}~\bibnamefont
  {Knetter}}\ and\ \bibinfo {author} {\bibfnamefont {G.}~\bibnamefont
  {Uhrig}},\ }\href {\doibase 10.1007/s100510050026} {\bibfield  {journal}
  {\bibinfo  {journal} {The European Physical Journal B}\ }\textbf {\bibinfo
  {volume} {13}},\ \bibinfo {pages} {209} (\bibinfo {year} {2000})}\BibitemShut
  {NoStop}%
\bibitem [{\citenamefont {Zheng}\ \emph {et~al.}(2001)\citenamefont {Zheng},
  \citenamefont {Hamer}, \citenamefont {Singh}, \citenamefont {Trebst},\ and\
  \citenamefont {Monien}}]{zheng01b}%
  \BibitemOpen
  \bibfield  {author} {\bibinfo {author} {\bibfnamefont {W.}~\bibnamefont
  {Zheng}}, \bibinfo {author} {\bibfnamefont {C.~J.}\ \bibnamefont {Hamer}},
  \bibinfo {author} {\bibfnamefont {R.~R.~P.}\ \bibnamefont {Singh}}, \bibinfo
  {author} {\bibfnamefont {S.}~\bibnamefont {Trebst}}, \ and\ \bibinfo {author}
  {\bibfnamefont {H.}~\bibnamefont {Monien}},\ }\href@noop {} {\bibfield
  {journal} {\bibinfo  {journal} {Phys. Rev. B}\ }\textbf {\bibinfo {volume}
  {63}},\ \bibinfo {pages} {144411} (\bibinfo {year} {2001})}\BibitemShut
  {NoStop}%
\bibitem [{\citenamefont {Schmidt}\ and\ \citenamefont
  {Uhrig}(2003)}]{Schmidt03}%
  \BibitemOpen
  \bibfield  {author} {\bibinfo {author} {\bibfnamefont {K.~P.}\ \bibnamefont
  {Schmidt}}\ and\ \bibinfo {author} {\bibfnamefont {G.~S.}\ \bibnamefont
  {Uhrig}},\ }\href {\doibase 10.1103/PhysRevLett.90.227204} {\bibfield
  {journal} {\bibinfo  {journal} {Phys. Rev. Lett.}\ }\textbf {\bibinfo
  {volume} {90}},\ \bibinfo {pages} {227204} (\bibinfo {year}
  {2003})}\BibitemShut {NoStop}%
\bibitem [{\citenamefont {Papenbrock}\ \emph {et~al.}(2003)\citenamefont
  {Papenbrock}, \citenamefont {Barnes}, \citenamefont {Dean}, \citenamefont
  {Stoitsov},\ and\ \citenamefont {Strayer}}]{papen03}%
  \BibitemOpen
  \bibfield  {author} {\bibinfo {author} {\bibfnamefont {T.}~\bibnamefont
  {Papenbrock}}, \bibinfo {author} {\bibfnamefont {T.}~\bibnamefont {Barnes}},
  \bibinfo {author} {\bibfnamefont {D.~J.}\ \bibnamefont {Dean}}, \bibinfo
  {author} {\bibfnamefont {M.~V.}\ \bibnamefont {Stoitsov}}, \ and\ \bibinfo
  {author} {\bibfnamefont {M.~R.}\ \bibnamefont {Strayer}},\ }\href@noop {}
  {\bibfield  {journal} {\bibinfo  {journal} {Phys. Rev. B}\ }\textbf {\bibinfo
  {volume} {68}},\ \bibinfo {pages} {024416} (\bibinfo {year}
  {2003})}\BibitemShut {NoStop}%
\bibitem [{\citenamefont {Auerbach}(1994)}]{auerb94}%
  \BibitemOpen
  \bibfield  {author} {\bibinfo {author} {\bibfnamefont {A.}~\bibnamefont
  {Auerbach}},\ }\href@noop {} {\emph {\bibinfo {title} {Interacting Electrons
  and Quantum Magnetism}}},\ Graduate Texts in Contemporary Physics\ (\bibinfo
  {publisher} {Springer},\ \bibinfo {address} {New York},\ \bibinfo {year}
  {1994})\BibitemShut {NoStop}%
\bibitem [{\citenamefont {Balents}(2010)}]{balen10}%
  \BibitemOpen
  \bibfield  {author} {\bibinfo {author} {\bibfnamefont {L.}~\bibnamefont
  {Balents}},\ }\href@noop {} {\bibfield  {journal} {\bibinfo  {journal}
  {Nature}\ }\textbf {\bibinfo {volume} {464}},\ \bibinfo {pages} {199}
  (\bibinfo {year} {2010})}\BibitemShut {NoStop}%
\bibitem [{\citenamefont {Fabrizio}\ \emph {et~al.}(1999)\citenamefont
  {Fabrizio}, \citenamefont {Gogolin},\ and\ \citenamefont
  {Nersesyan}}]{Fabrizio99}%
  \BibitemOpen
  \bibfield  {author} {\bibinfo {author} {\bibfnamefont {M.}~\bibnamefont
  {Fabrizio}}, \bibinfo {author} {\bibfnamefont {A.~O.}\ \bibnamefont
  {Gogolin}}, \ and\ \bibinfo {author} {\bibfnamefont {A.~A.}\ \bibnamefont
  {Nersesyan}},\ }\href {\doibase 10.1103/PhysRevLett.83.2014} {\bibfield
  {journal} {\bibinfo  {journal} {Phys. Rev. Lett.}\ }\textbf {\bibinfo
  {volume} {83}},\ \bibinfo {pages} {2014} (\bibinfo {year}
  {1999})}\BibitemShut {NoStop}%
\bibitem [{\citenamefont {Torio}\ \emph {et~al.}(2001)\citenamefont {Torio},
  \citenamefont {Aligia},\ and\ \citenamefont {Ceccatto}}]{Torio01}%
  \BibitemOpen
  \bibfield  {author} {\bibinfo {author} {\bibfnamefont {M.~E.}\ \bibnamefont
  {Torio}}, \bibinfo {author} {\bibfnamefont {A.~A.}\ \bibnamefont {Aligia}}, \
  and\ \bibinfo {author} {\bibfnamefont {H.~A.}\ \bibnamefont {Ceccatto}},\
  }\href {\doibase 10.1103/PhysRevB.64.121105} {\bibfield  {journal} {\bibinfo
  {journal} {Phys. Rev. B}\ }\textbf {\bibinfo {volume} {64}},\ \bibinfo
  {pages} {121105} (\bibinfo {year} {2001})}\BibitemShut {NoStop}%
\bibitem [{\citenamefont {Manmana}\ \emph {et~al.}(2004)\citenamefont
  {Manmana}, \citenamefont {Meden}, \citenamefont {Noack},\ and\ \citenamefont
  {Sch\"onhammer}}]{Manmana04}%
  \BibitemOpen
  \bibfield  {author} {\bibinfo {author} {\bibfnamefont {S.~R.}\ \bibnamefont
  {Manmana}}, \bibinfo {author} {\bibfnamefont {V.}~\bibnamefont {Meden}},
  \bibinfo {author} {\bibfnamefont {R.~M.}\ \bibnamefont {Noack}}, \ and\
  \bibinfo {author} {\bibfnamefont {K.}~\bibnamefont {Sch\"onhammer}},\ }\href
  {\doibase 10.1103/PhysRevB.70.155115} {\bibfield  {journal} {\bibinfo
  {journal} {Phys. Rev. B}\ }\textbf {\bibinfo {volume} {70}},\ \bibinfo
  {pages} {155115} (\bibinfo {year} {2004})}\BibitemShut {NoStop}%
\bibitem [{\citenamefont {Otsuka}\ and\ \citenamefont
  {Nakamura}(2005)}]{Otsuka2005}%
  \BibitemOpen
  \bibfield  {author} {\bibinfo {author} {\bibfnamefont {H.}~\bibnamefont
  {Otsuka}}\ and\ \bibinfo {author} {\bibfnamefont {M.}~\bibnamefont
  {Nakamura}},\ }\href {\doibase 10.1103/PhysRevB.71.155105} {\bibfield
  {journal} {\bibinfo  {journal} {Phys. Rev. B}\ }\textbf {\bibinfo {volume}
  {71}},\ \bibinfo {pages} {155105} (\bibinfo {year} {2005})}\BibitemShut
  {NoStop}%
\bibitem [{\citenamefont {Tincani}\ \emph {et~al.}(2009)\citenamefont
  {Tincani}, \citenamefont {Noack},\ and\ \citenamefont
  {Baeriswyl}}]{Tincani09}%
  \BibitemOpen
  \bibfield  {author} {\bibinfo {author} {\bibfnamefont {L.}~\bibnamefont
  {Tincani}}, \bibinfo {author} {\bibfnamefont {R.~M.}\ \bibnamefont {Noack}},
  \ and\ \bibinfo {author} {\bibfnamefont {D.}~\bibnamefont {Baeriswyl}},\
  }\href {\doibase 10.1103/PhysRevB.79.165109} {\bibfield  {journal} {\bibinfo
  {journal} {Phys. Rev. B}\ }\textbf {\bibinfo {volume} {79}},\ \bibinfo
  {pages} {165109} (\bibinfo {year} {2009})}\BibitemShut {NoStop}%
\bibitem [{\citenamefont {Hafez~Torbati}\ \emph {et~al.}(2014)\citenamefont
  {Hafez~Torbati}, \citenamefont {Drescher},\ and\ \citenamefont
  {Uhrig}}]{HafezTorbati2014}%
  \BibitemOpen
  \bibfield  {author} {\bibinfo {author} {\bibfnamefont {M.}~\bibnamefont
  {Hafez~Torbati}}, \bibinfo {author} {\bibfnamefont {N.~A.}\ \bibnamefont
  {Drescher}}, \ and\ \bibinfo {author} {\bibfnamefont {G.~S.}\ \bibnamefont
  {Uhrig}},\ }\href {\doibase 10.1103/PhysRevB.89.245126} {\bibfield  {journal}
  {\bibinfo  {journal} {Phys. Rev. B}\ }\textbf {\bibinfo {volume} {89}},\
  \bibinfo {pages} {245126} (\bibinfo {year} {2014})}\BibitemShut {NoStop}%
\bibitem [{\citenamefont {Garg}\ \emph {et~al.}(2006)\citenamefont {Garg},
  \citenamefont {Krishnamurthy},\ and\ \citenamefont {Randeria}}]{Garg06}%
  \BibitemOpen
  \bibfield  {author} {\bibinfo {author} {\bibfnamefont {A.}~\bibnamefont
  {Garg}}, \bibinfo {author} {\bibfnamefont {H.~R.}\ \bibnamefont
  {Krishnamurthy}}, \ and\ \bibinfo {author} {\bibfnamefont {M.}~\bibnamefont
  {Randeria}},\ }\href {\doibase 10.1103/PhysRevLett.97.046403} {\bibfield
  {journal} {\bibinfo  {journal} {Phys. Rev. Lett.}\ }\textbf {\bibinfo
  {volume} {97}},\ \bibinfo {pages} {046403} (\bibinfo {year}
  {2006})}\BibitemShut {NoStop}%
\bibitem [{\citenamefont {Kancharla}\ and\ \citenamefont
  {Dagotto}(2007)}]{Kancharla07}%
  \BibitemOpen
  \bibfield  {author} {\bibinfo {author} {\bibfnamefont {S.~S.}\ \bibnamefont
  {Kancharla}}\ and\ \bibinfo {author} {\bibfnamefont {E.}~\bibnamefont
  {Dagotto}},\ }\href {\doibase 10.1103/PhysRevLett.98.016402} {\bibfield
  {journal} {\bibinfo  {journal} {Phys. Rev. Lett.}\ }\textbf {\bibinfo
  {volume} {98}},\ \bibinfo {pages} {016402} (\bibinfo {year}
  {2007})}\BibitemShut {NoStop}%
\bibitem [{\citenamefont {Paris}\ \emph {et~al.}(2007)\citenamefont {Paris},
  \citenamefont {Bouadim}, \citenamefont {H\'ebert}, \citenamefont {Batrouni},\
  and\ \citenamefont {Scalettar}}]{Paris07}%
  \BibitemOpen
  \bibfield  {author} {\bibinfo {author} {\bibfnamefont {N.}~\bibnamefont
  {Paris}}, \bibinfo {author} {\bibfnamefont {K.}~\bibnamefont {Bouadim}},
  \bibinfo {author} {\bibfnamefont {F.}~\bibnamefont {H\'ebert}}, \bibinfo
  {author} {\bibfnamefont {G.~G.}\ \bibnamefont {Batrouni}}, \ and\ \bibinfo
  {author} {\bibfnamefont {R.~T.}\ \bibnamefont {Scalettar}},\ }\href {\doibase
  10.1103/PhysRevLett.98.046403} {\bibfield  {journal} {\bibinfo  {journal}
  {Phys. Rev. Lett.}\ }\textbf {\bibinfo {volume} {98}},\ \bibinfo {pages}
  {046403} (\bibinfo {year} {2007})}\BibitemShut {NoStop}%
\bibitem [{\citenamefont {Craco}\ \emph {et~al.}(2008)\citenamefont {Craco},
  \citenamefont {Lombardo}, \citenamefont {Hayn}, \citenamefont {Japaridze},\
  and\ \citenamefont {M\"uller-Hartmann}}]{Craco08}%
  \BibitemOpen
  \bibfield  {author} {\bibinfo {author} {\bibfnamefont {L.}~\bibnamefont
  {Craco}}, \bibinfo {author} {\bibfnamefont {P.}~\bibnamefont {Lombardo}},
  \bibinfo {author} {\bibfnamefont {R.}~\bibnamefont {Hayn}}, \bibinfo {author}
  {\bibfnamefont {G.~I.}\ \bibnamefont {Japaridze}}, \ and\ \bibinfo {author}
  {\bibfnamefont {E.}~\bibnamefont {M\"uller-Hartmann}},\ }\href {\doibase
  10.1103/PhysRevB.78.075121} {\bibfield  {journal} {\bibinfo  {journal} {Phys.
  Rev. B}\ }\textbf {\bibinfo {volume} {78}},\ \bibinfo {pages} {075121}
  (\bibinfo {year} {2008})}\BibitemShut {NoStop}%
\bibitem [{\citenamefont {Chen}\ \emph {et~al.}(2010)\citenamefont {Chen},
  \citenamefont {Zhao}, \citenamefont {Lin},\ and\ \citenamefont
  {Wu}}]{Chen10}%
  \BibitemOpen
  \bibfield  {author} {\bibinfo {author} {\bibfnamefont {H.-M.}\ \bibnamefont
  {Chen}}, \bibinfo {author} {\bibfnamefont {H.}~\bibnamefont {Zhao}}, \bibinfo
  {author} {\bibfnamefont {H.-Q.}\ \bibnamefont {Lin}}, \ and\ \bibinfo
  {author} {\bibfnamefont {C.-Q.}\ \bibnamefont {Wu}},\ }\href
  {http://stacks.iop.org/1367-2630/12/i=9/a=093021} {\bibfield  {journal}
  {\bibinfo  {journal} {New Journal of Physics}\ }\textbf {\bibinfo {volume}
  {12}},\ \bibinfo {pages} {093021} (\bibinfo {year} {2010})}\BibitemShut
  {NoStop}%
\bibitem [{\citenamefont {Strebel}\ and\ \citenamefont
  {Soos}(1970)}]{Strebel1970}%
  \BibitemOpen
  \bibfield  {author} {\bibinfo {author} {\bibfnamefont {P.~J.}\ \bibnamefont
  {Strebel}}\ and\ \bibinfo {author} {\bibfnamefont {Z.~G.}\ \bibnamefont
  {Soos}},\ }\href {\doibase http://dx.doi.org/10.1063/1.1673881} {\bibfield
  {journal} {\bibinfo  {journal} {The Journal of Chemical Physics}\ }\textbf
  {\bibinfo {volume} {53}},\ \bibinfo {pages} {4077} (\bibinfo {year}
  {1970})}\BibitemShut {NoStop}%
\bibitem [{\citenamefont {Soos}\ and\ \citenamefont
  {Mazumdar}(1978)}]{Soos1978}%
  \BibitemOpen
  \bibfield  {author} {\bibinfo {author} {\bibfnamefont {Z.~G.}\ \bibnamefont
  {Soos}}\ and\ \bibinfo {author} {\bibfnamefont {S.}~\bibnamefont
  {Mazumdar}},\ }\href {\doibase 10.1103/PhysRevB.18.1991} {\bibfield
  {journal} {\bibinfo  {journal} {Phys. Rev. B}\ }\textbf {\bibinfo {volume}
  {18}},\ \bibinfo {pages} {1991} (\bibinfo {year} {1978})}\BibitemShut
  {NoStop}%
\bibitem [{\citenamefont {Nagaosa}\ and\ \citenamefont
  {Takimoto}(1986)}]{Nagaosa86a}%
  \BibitemOpen
  \bibfield  {author} {\bibinfo {author} {\bibfnamefont {N.}~\bibnamefont
  {Nagaosa}}\ and\ \bibinfo {author} {\bibfnamefont {J.}~\bibnamefont
  {Takimoto}},\ }\href {\doibase 10.1143/JPSJ.55.2735} {\bibfield  {journal}
  {\bibinfo  {journal} {Journal of the Physical Society of Japan}\ }\textbf
  {\bibinfo {volume} {55}},\ \bibinfo {pages} {2735} (\bibinfo {year}
  {1986})}\BibitemShut {NoStop}%
\bibitem [{\citenamefont {Torrance}\ \emph {et~al.}(1981)\citenamefont
  {Torrance}, \citenamefont {Vazquez}, \citenamefont {Mayerle},\ and\
  \citenamefont {Lee}}]{Torrance81a}%
  \BibitemOpen
  \bibfield  {author} {\bibinfo {author} {\bibfnamefont {J.~B.}\ \bibnamefont
  {Torrance}}, \bibinfo {author} {\bibfnamefont {J.~E.}\ \bibnamefont
  {Vazquez}}, \bibinfo {author} {\bibfnamefont {J.~J.}\ \bibnamefont
  {Mayerle}}, \ and\ \bibinfo {author} {\bibfnamefont {V.~Y.}\ \bibnamefont
  {Lee}},\ }\href {\doibase 10.1103/PhysRevLett.46.253} {\bibfield  {journal}
  {\bibinfo  {journal} {Phys. Rev. Lett.}\ }\textbf {\bibinfo {volume} {46}},\
  \bibinfo {pages} {253} (\bibinfo {year} {1981})}\BibitemShut {NoStop}%
\bibitem [{\citenamefont {Egami}\ \emph {et~al.}(1993)\citenamefont {Egami},
  \citenamefont {Ishihara},\ and\ \citenamefont {Tachiki}}]{Egami93}%
  \BibitemOpen
  \bibfield  {author} {\bibinfo {author} {\bibfnamefont {T.}~\bibnamefont
  {Egami}}, \bibinfo {author} {\bibfnamefont {S.}~\bibnamefont {Ishihara}}, \
  and\ \bibinfo {author} {\bibfnamefont {M.}~\bibnamefont {Tachiki}},\ }\href
  {\doibase 10.1126/science.261.5126.1307} {\bibfield  {journal} {\bibinfo
  {journal} {Science}\ }\textbf {\bibinfo {volume} {261}},\ \bibinfo {pages}
  {1307} (\bibinfo {year} {1993})}\BibitemShut {NoStop}%
\bibitem [{\citenamefont {Kobayashi}\ \emph {et~al.}(2012)\citenamefont
  {Kobayashi}, \citenamefont {Horiuchi}, \citenamefont {Kumai}, \citenamefont
  {Kagawa}, \citenamefont {Murakami},\ and\ \citenamefont
  {Tokura}}]{Kobayashi12}%
  \BibitemOpen
  \bibfield  {author} {\bibinfo {author} {\bibfnamefont {K.}~\bibnamefont
  {Kobayashi}}, \bibinfo {author} {\bibfnamefont {S.}~\bibnamefont {Horiuchi}},
  \bibinfo {author} {\bibfnamefont {R.}~\bibnamefont {Kumai}}, \bibinfo
  {author} {\bibfnamefont {F.}~\bibnamefont {Kagawa}}, \bibinfo {author}
  {\bibfnamefont {Y.}~\bibnamefont {Murakami}}, \ and\ \bibinfo {author}
  {\bibfnamefont {Y.}~\bibnamefont {Tokura}},\ }\href {\doibase
  10.1103/PhysRevLett.108.237601} {\bibfield  {journal} {\bibinfo  {journal}
  {Phys. Rev. Lett.}\ }\textbf {\bibinfo {volume} {108}},\ \bibinfo {pages}
  {237601} (\bibinfo {year} {2012})}\BibitemShut {NoStop}%
\bibitem [{\citenamefont {Krull}\ \emph {et~al.}(2012)\citenamefont {Krull},
  \citenamefont {Drescher},\ and\ \citenamefont {Uhrig}}]{Krull12}%
  \BibitemOpen
  \bibfield  {author} {\bibinfo {author} {\bibfnamefont {H.}~\bibnamefont
  {Krull}}, \bibinfo {author} {\bibfnamefont {N.~A.}\ \bibnamefont {Drescher}},
  \ and\ \bibinfo {author} {\bibfnamefont {G.~S.}\ \bibnamefont {Uhrig}},\
  }\href {\doibase 10.1103/PhysRevB.86.125113} {\bibfield  {journal} {\bibinfo
  {journal} {Phys. Rev. B}\ }\textbf {\bibinfo {volume} {86}},\ \bibinfo
  {pages} {125113} (\bibinfo {year} {2012})}\BibitemShut {NoStop}%
\bibitem [{\citenamefont {Hafez}\ and\ \citenamefont
  {Jafari}(2010)}]{Hafez10b}%
  \BibitemOpen
  \bibfield  {author} {\bibinfo {author} {\bibfnamefont {M.}~\bibnamefont
  {Hafez}}\ and\ \bibinfo {author} {\bibfnamefont {S.~A.}\ \bibnamefont
  {Jafari}},\ }\href {\doibase 10.1140/epjb/e2010-10509-x} {\bibfield
  {journal} {\bibinfo  {journal} {The European Physical Journal B}\ }\textbf
  {\bibinfo {volume} {78}},\ \bibinfo {pages} {323} (\bibinfo {year}
  {2010})}\BibitemShut {NoStop}%
\bibitem [{\citenamefont {Hafez}\ and\ \citenamefont
  {Abolhassani}(2011)}]{Hafez11}%
  \BibitemOpen
  \bibfield  {author} {\bibinfo {author} {\bibfnamefont {M.}~\bibnamefont
  {Hafez}}\ and\ \bibinfo {author} {\bibfnamefont {M.~R.}\ \bibnamefont
  {Abolhassani}},\ }\href {http://stacks.iop.org/0953-8984/23/i=24/a=245602}
  {\bibfield  {journal} {\bibinfo  {journal} {Journal of Physics: Condensed
  Matter}\ }\textbf {\bibinfo {volume} {23}},\ \bibinfo {pages} {245602}
  (\bibinfo {year} {2011})}\BibitemShut {NoStop}%
\bibitem [{\citenamefont {Oitmaa}\ \emph {et~al.}(2006)\citenamefont {Oitmaa},
  \citenamefont {Hamer},\ and\ \citenamefont {Zheng}}]{Oitmaa06}%
  \BibitemOpen
  \bibfield  {author} {\bibinfo {author} {\bibfnamefont {J.}~\bibnamefont
  {Oitmaa}}, \bibinfo {author} {\bibfnamefont {C.}~\bibnamefont {Hamer}}, \
  and\ \bibinfo {author} {\bibfnamefont {W.}~\bibnamefont {Zheng}},\
  }\href@noop {} {\emph {\bibinfo {title} {Series Expansion Methods for
  Strongly Interacting Lattice Models}}}\ (\bibinfo  {publisher} {Cambridge
  University Press},\ \bibinfo {address} {Cambridge},\ \bibinfo {year}
  {2006})\BibitemShut {NoStop}%
\bibitem [{\citenamefont {Duffe}\ and\ \citenamefont {Uhrig}(2011)}]{Duffe11}%
  \BibitemOpen
  \bibfield  {author} {\bibinfo {author} {\bibfnamefont {S.}~\bibnamefont
  {Duffe}}\ and\ \bibinfo {author} {\bibfnamefont {G.}~\bibnamefont {Uhrig}},\
  }\href {\doibase 10.1140/epjb/e2011-20150-x} {\bibfield  {journal} {\bibinfo
  {journal} {The European Physical Journal B}\ }\textbf {\bibinfo {volume}
  {84}},\ \bibinfo {pages} {475} (\bibinfo {year} {2011})}\BibitemShut
  {NoStop}%
\bibitem [{\citenamefont {Wegner}(1994)}]{Wegner94}%
  \BibitemOpen
  \bibfield  {author} {\bibinfo {author} {\bibfnamefont {F.}~\bibnamefont
  {Wegner}},\ }\href {\doibase 10.1002/andp.19945060203} {\bibfield  {journal}
  {\bibinfo  {journal} {Annalen der Physik}\ }\textbf {\bibinfo {volume}
  {506}},\ \bibinfo {pages} {77} (\bibinfo {year} {1994})}\BibitemShut
  {NoStop}%
\bibitem [{\citenamefont {Kehrein}(2006)}]{Kehrein06}%
  \BibitemOpen
  \bibfield  {author} {\bibinfo {author} {\bibfnamefont {S.}~\bibnamefont
  {Kehrein}},\ }\href
  {http://www.springer.com/materials/book/978-3-540-34067-6} {\emph {\bibinfo
  {title} {The Flow Equation Approach to Many-Particle Systems}}},\ Springer
  Tracts in Modern Physics, Vol. 217\ (\bibinfo  {publisher} {Springer},\
  \bibinfo {address} {Berlin},\ \bibinfo {year} {2006})\BibitemShut {NoStop}%
\bibitem [{\citenamefont {Knetter}\ \emph {et~al.}(2003)\citenamefont
  {Knetter}, \citenamefont {Schmidt},\ and\ \citenamefont
  {Uhrig}}]{Knetter03a}%
  \BibitemOpen
  \bibfield  {author} {\bibinfo {author} {\bibfnamefont {C.}~\bibnamefont
  {Knetter}}, \bibinfo {author} {\bibfnamefont {K.}~\bibnamefont {Schmidt}}, \
  and\ \bibinfo {author} {\bibfnamefont {G.}~\bibnamefont {Uhrig}},\ }\href
  {\doibase 10.1140/epjb/e2004-00008-2} {\bibfield  {journal} {\bibinfo
  {journal} {The European Physical Journal B}\ }\textbf {\bibinfo {volume}
  {36}},\ \bibinfo {pages} {525} (\bibinfo {year} {2003})}\BibitemShut
  {NoStop}%
\bibitem [{\citenamefont {Yang}\ \emph {et~al.}(2010)\citenamefont {Yang},
  \citenamefont {L\"auchli}, \citenamefont {Mila},\ and\ \citenamefont
  {Schmidt}}]{Yang10}%
  \BibitemOpen
  \bibfield  {author} {\bibinfo {author} {\bibfnamefont {H.-Y.}\ \bibnamefont
  {Yang}}, \bibinfo {author} {\bibfnamefont {A.~M.}\ \bibnamefont {L\"auchli}},
  \bibinfo {author} {\bibfnamefont {F.}~\bibnamefont {Mila}}, \ and\ \bibinfo
  {author} {\bibfnamefont {K.~P.}\ \bibnamefont {Schmidt}},\ }\href@noop {}
  {\bibfield  {journal} {\bibinfo  {journal} {Phys. Rev. Lett.}\ }\textbf
  {\bibinfo {volume} {105}},\ \bibinfo {pages} {267204} (\bibinfo {year}
  {2010})}\BibitemShut {NoStop}%
\bibitem [{\citenamefont {Dusuel}\ and\ \citenamefont
  {Uhrig}(2004)}]{Dusuel04}%
  \BibitemOpen
  \bibfield  {author} {\bibinfo {author} {\bibfnamefont {S.}~\bibnamefont
  {Dusuel}}\ and\ \bibinfo {author} {\bibfnamefont {G.~S.}\ \bibnamefont
  {Uhrig}},\ }\href {http://stacks.iop.org/0305-4470/37/i=39/a=014} {\bibfield
  {journal} {\bibinfo  {journal} {Journal of Physics A: Mathematical and
  General}\ }\textbf {\bibinfo {volume} {37}},\ \bibinfo {pages} {9275}
  (\bibinfo {year} {2004})}\BibitemShut {NoStop}%
\bibitem [{\citenamefont {Yang}\ and\ \citenamefont {Schmidt}(2011)}]{Yang11}%
  \BibitemOpen
  \bibfield  {author} {\bibinfo {author} {\bibfnamefont {H.~Y.}\ \bibnamefont
  {Yang}}\ and\ \bibinfo {author} {\bibfnamefont {K.~P.}\ \bibnamefont
  {Schmidt}},\ }\href {http://stacks.iop.org/0295-5075/94/i=1/a=17004}
  {\bibfield  {journal} {\bibinfo  {journal} {EPL (Europhysics Letters)}\
  }\textbf {\bibinfo {volume} {94}},\ \bibinfo {pages} {17004} (\bibinfo {year}
  {2011})}\BibitemShut {NoStop}%
\bibitem [{\citenamefont {Fauseweh}\ and\ \citenamefont
  {Uhrig}(2013)}]{Fauseweh13}%
  \BibitemOpen
  \bibfield  {author} {\bibinfo {author} {\bibfnamefont {B.}~\bibnamefont
  {Fauseweh}}\ and\ \bibinfo {author} {\bibfnamefont {G.~S.}\ \bibnamefont
  {Uhrig}},\ }\href {\doibase 10.1103/PhysRevB.87.184406} {\bibfield  {journal}
  {\bibinfo  {journal} {Phys. Rev. B}\ }\textbf {\bibinfo {volume} {87}},\
  \bibinfo {pages} {184406} (\bibinfo {year} {2013})}\BibitemShut {NoStop}%
\bibitem [{\citenamefont {Fischer}\ \emph {et~al.}(2010)\citenamefont
  {Fischer}, \citenamefont {Duffe},\ and\ \citenamefont {Uhrig}}]{Fischer10}%
  \BibitemOpen
  \bibfield  {author} {\bibinfo {author} {\bibfnamefont {T.}~\bibnamefont
  {Fischer}}, \bibinfo {author} {\bibfnamefont {S.}~\bibnamefont {Duffe}}, \
  and\ \bibinfo {author} {\bibfnamefont {G.~S.}\ \bibnamefont {Uhrig}},\ }\href
  {http://stacks.iop.org/1367-2630/12/i=3/a=033048} {\bibfield  {journal}
  {\bibinfo  {journal} {New Journal of Physics}\ }\textbf {\bibinfo {volume}
  {12}},\ \bibinfo {pages} {033048} (\bibinfo {year} {2010})}\BibitemShut
  {NoStop}%
\bibitem [{\citenamefont {M\"uller}\ \emph {et~al.}(1981)\citenamefont
  {M\"uller}, \citenamefont {Thomas}, \citenamefont {Beck},\ and\ \citenamefont
  {Bonner}}]{mulle81}%
  \BibitemOpen
  \bibfield  {author} {\bibinfo {author} {\bibfnamefont {G.}~\bibnamefont
  {M\"uller}}, \bibinfo {author} {\bibfnamefont {H.}~\bibnamefont {Thomas}},
  \bibinfo {author} {\bibfnamefont {H.}~\bibnamefont {Beck}}, \ and\ \bibinfo
  {author} {\bibfnamefont {J.~C.}\ \bibnamefont {Bonner}},\ }\href@noop {}
  {\bibfield  {journal} {\bibinfo  {journal} {Phys. Rev. B}\ }\textbf {\bibinfo
  {volume} {24}},\ \bibinfo {pages} {1429} (\bibinfo {year}
  {1981})}\BibitemShut {NoStop}%
\bibitem [{\citenamefont {Karbach}\ \emph {et~al.}(1997)\citenamefont
  {Karbach}, \citenamefont {M\"uller}, \citenamefont {Bougourzi}, \citenamefont
  {Fledderjohann},\ and\ \citenamefont {M\"utter}}]{Karbach97}%
  \BibitemOpen
  \bibfield  {author} {\bibinfo {author} {\bibfnamefont {M.}~\bibnamefont
  {Karbach}}, \bibinfo {author} {\bibfnamefont {G.}~\bibnamefont {M\"uller}},
  \bibinfo {author} {\bibfnamefont {A.~H.}\ \bibnamefont {Bougourzi}}, \bibinfo
  {author} {\bibfnamefont {A.}~\bibnamefont {Fledderjohann}}, \ and\ \bibinfo
  {author} {\bibfnamefont {K.-H.}\ \bibnamefont {M\"utter}},\ }\href {\doibase
  10.1103/PhysRevB.55.12510} {\bibfield  {journal} {\bibinfo  {journal} {Phys.
  Rev. B}\ }\textbf {\bibinfo {volume} {55}},\ \bibinfo {pages} {12510}
  (\bibinfo {year} {1997})}\BibitemShut {NoStop}%
\bibitem [{\citenamefont {Gogolin}\ \emph {et~al.}(1998)\citenamefont
  {Gogolin}, \citenamefont {Nersesyan},\ and\ \citenamefont
  {Tsvelik}}]{gogol98}%
  \BibitemOpen
  \bibfield  {author} {\bibinfo {author} {\bibfnamefont {A.~O.}\ \bibnamefont
  {Gogolin}}, \bibinfo {author} {\bibfnamefont {A.~A.}\ \bibnamefont
  {Nersesyan}}, \ and\ \bibinfo {author} {\bibfnamefont {A.~M.}\ \bibnamefont
  {Tsvelik}},\ }\href@noop {} {\emph {\bibinfo {title} {Bosonization and
  Strongly Correlated Systems}}}\ (\bibinfo  {publisher} {Cambridge University
  Press},\ \bibinfo {address} {Cambridge},\ \bibinfo {year} {1998})\BibitemShut
  {NoStop}%
\bibitem [{\citenamefont {Wilkens}\ and\ \citenamefont
  {Martin}(2001)}]{Wilkens01}%
  \BibitemOpen
  \bibfield  {author} {\bibinfo {author} {\bibfnamefont {T.}~\bibnamefont
  {Wilkens}}\ and\ \bibinfo {author} {\bibfnamefont {R.~M.}\ \bibnamefont
  {Martin}},\ }\href {\doibase 10.1103/PhysRevB.63.235108} {\bibfield
  {journal} {\bibinfo  {journal} {Phys. Rev. B}\ }\textbf {\bibinfo {volume}
  {63}},\ \bibinfo {pages} {235108} (\bibinfo {year} {2001})}\BibitemShut
  {NoStop}%
\bibitem [{\citenamefont {Kampf}\ \emph {et~al.}(2003)\citenamefont {Kampf},
  \citenamefont {Sekania}, \citenamefont {Japaridze},\ and\ \citenamefont
  {Brune}}]{Kampf03}%
  \BibitemOpen
  \bibfield  {author} {\bibinfo {author} {\bibfnamefont {A.~P.}\ \bibnamefont
  {Kampf}}, \bibinfo {author} {\bibfnamefont {M.}~\bibnamefont {Sekania}},
  \bibinfo {author} {\bibfnamefont {G.~I.}\ \bibnamefont {Japaridze}}, \ and\
  \bibinfo {author} {\bibfnamefont {P.}~\bibnamefont {Brune}},\ }\href
  {http://stacks.iop.org/0953-8984/15/i=34/a=319} {\bibfield  {journal}
  {\bibinfo  {journal} {Journal of Physics: Condensed Matter}\ }\textbf
  {\bibinfo {volume} {15}},\ \bibinfo {pages} {5895} (\bibinfo {year}
  {2003})}\BibitemShut {NoStop}%
\end{thebibliography}


\end{document}